\documentclass{article}

\usepackage{arxiv}

\usepackage[utf8]{inputenc}
\usepackage[T1]{fontenc}
\usepackage{url}
\usepackage{booktabs}
\usepackage{amsmath}
\usepackage{amsfonts}
\usepackage{amssymb}
\usepackage{nicefrac}
\usepackage{microtype}
\usepackage{graphicx}
\usepackage[svgnames]{xcolor}
\usepackage[colorlinks=true,linkcolor=NavyBlue,citecolor=NavyBlue,urlcolor=NavyBlue]{hyperref}
\usepackage{subcaption}
\usepackage{array}
\usepackage{multirow}
\usepackage{siunitx}
\usepackage{makecell}
\usepackage[font=small]{caption}
\usepackage[section]{placeins}

\usepackage[authoryear,round]{natbib}
\setcitestyle{aysep={}}

\graphicspath{{./images/}}
\title{Ultra-Compact CNN Architectures for Tropical Bird Audio Detection on Microcontrollers}

\author{
  Muhammad Mun'im Ahmad Zabidi$^{1,2}$ \quad
  Mohd Yamani Idna Idris$^{1}$ \quad
  Norisma Idris$^{1}$ \\[4pt]
  $^{1}$Faculty of Computer Science and Information Technology,
    Universiti Malaya, Kuala Lumpur, Malaysia \\
  $^{2}$Faculty of Electrical Engineering, Universiti Teknologi Malaysia,
    Johor Bahru, Malaysia \\[4pt]
  \texttt{yamani@um.edu.my}
}

\begin{document}
\maketitle
\begin{abstract}
Passive acoustic monitoring of tropical biodiversity is bottlenecked by the storage and battery cost of continuously recording soundscapes in which bird vocalisations typically occupy less than 10\% of the audio. Autonomous recording units built on low-power microcontrollers (typically ARM Cortex-M with ${\leq}256$\,kB of RAM) address this by triggering only on likely-positive segments, but the on-device options are unsatisfying: coarse frequency-energy triggers such as Goertzel filters flood SD cards with false positives at ${\approx}71\%$ precision, whereas neural detectors developed for temperate single-species tasks are either too large to deploy or transfer poorly to species-rich tropical settings. We present DrongoNet, a family of three INT8 CNN detectors sized for this envelope and validated on a 50{,}000-clip, 1{,}677-species Southeast Asian tropical dataset (SEABAD). The headline model, \textbf{DrongoNet-Micro} (919 parameters, 6.26\,kB, $0.9810$ AUC, $98.3\%$ mean recall at $\tau{=}0.35$), is a drop-in replacement for the Goertzel trigger used in commodity field recorders: at $\alpha{=}0.10$ tropical prevalence it captures 8\,pp more bird vocalisations than Goertzel and extends a 32\,GB card from ${\approx}28$ to ${\approx}45$ days of monitoring. \textbf{DrongoNet-Nano} (5.09\,kB) bounds the ultra-low-flash extreme; \textbf{DrongoNet-Edge} (33.06\,kB, $0.9991$ AUC) targets Linux SBCs. On SEABAD, Micro matches a retrained TinyChirp CNN-Mel baseline within 0.1\,pp AUC at $28\times$ fewer parameters, confirming that the family is deployment-agnostic across mel-spectrogram bird corpora but requires per-environment retraining. Full INT8 quantisation costs ${<}0.12\%$ AUC across all three variants.
\end{abstract}
\vspace{1em}
\vspace{2em}

\section{Introduction}

Birds are widely used as bioindicators because they occupy diverse ecological niches, respond sensitively to habitat disturbance, and produce acoustically distinctive vocalisations. In tropical ecosystems, which contain more than half of the world's terrestrial biodiversity and are undergoing rapid environmental change, monitoring them at scale remains challenging. Passive acoustic monitoring (PAM) with autonomous recording units (ARUs) enables continuous, non-invasive surveys over large areas \citep{ross2023passive,xie2023review}, but bird vocalisations typically occupy only a small fraction of recorded audio.

In the strongly annotated Neotropical PteroSet corpus, only 9\% of 73.6 hours of recordings contain labelled bird vocalisations \citep{ruiz2026pteroset}. Huang et al. similarly reported that bird calls accounted for less than 10\% of recordings in a temperate single-species survey \citep{huang2024tinychirp}. Most recorded audio therefore consists of wind, rain, insects, and other non-target sounds.

Recording and storing these non-target segments increases storage use, consumes energy during SD-card writes, and adds to the manual effort required to review recordings. Commodity frequency-energy triggers, such as the Goertzel filter used by AudioMoth, reduce some of this overhead but remain too coarse to distinguish bird vocalisations reliably from background sounds. Bird Audio Detection (BAD), which determines whether an audio segment contains a bird vocalisation, provides a more effective first-stage filter.

Because BAD is a binary presence--absence task rather than species-level classification, it requires substantially fewer computational resources and is well suited to embedded deployment \citep{stowell2016bird,stowell2022computational,stowell2018automatic,kongkahing2025hornbill}. Recent advances in TinyML have also made neural inference practical on microcontroller-class hardware \citep{banbury2021micronets,david2021tensorflow,somvanshi2025tiny}, as reflected by the BioDCASE 2025 "bioacoustics on tiny hardware" challenge targeting ESP32-class devices \citep{carmantini2025biodcase}. However, it remains unclear which network architectures can achieve sufficiently high recall while satisfying the memory, energy, and computational constraints of these platforms.

A further challenge is the limited availability of tropical training data. Most publicly available bird sound datasets were collected in temperate regions \citep{kahl2021birdnet,stowell2022computational}, and models trained on them often generalise poorly to tropical soundscapes characterised by higher species richness, overlapping vocalisations, and prominent non-avian biophony. This limitation is particularly evident in Southeast Asia: a recent global evaluation of BirdNET across 67 recording sites reported a vocalisation-level recall of only 25.7\% in Asia, meaning that nearly three-quarters of bird vocalisations were missed, while dataset-level precision was only 27.4\%, with more than 72\% of reported species being false positives \citep{funosas2026global}.

These results highlight the need for detectors trained specifically for tropical environments. This study evaluates detector architectures using the recently introduced SEABAD (Southeast Asian Bird Audio Detection) dataset \citep{zabidi2026seabad}, which contains 50,000 curated three-second recordings covering 1,677 species. As the dataset is described elsewhere, it is used here solely as a benchmark for model evaluation and embedded deployment.

We introduce DrongoNet\footnote{Named for the drongo (family \textit{Dicruridae}), a vocally versatile Old World tropical passerine.}, a family of lightweight convolutional neural networks designed for tropical BAD under microcontroller-class resource constraints. Specifically, we investigate three questions: (1) whether architectures originally developed for single-species detection generalise to a multi-species tropical setting; (2) which architectural modifications provide the best accuracy--resource trade-offs at MCU scale; and (3) how detection thresholds should be calibrated when BAD is used as the front-end trigger for downstream species classification.

The contributions of this work are as follows:
\begin{itemize}
	\item We evaluate the TinyChirp architecture \citep{huang2024tinychirp} on SEABAD, showing that the published single-species model performs poorly in a zero-shot setting, whereas retraining on SEABAD yields a strong baseline for comparison.

	\item We conduct a four-phase ablation study examining frequency resolution, pooling strategies, loss functions, learnable frequency emphasis, and depthwise separable convolutions for MCU-class bird audio detection.

	\item We develop three DrongoNet variants spanning different resource budgets. After INT8 quantisation, the models occupy between 5.09\,kB and 33.06\,kB while achieving AUC values from 0.9727 to 0.9991.

	\item We characterise the complete embedded inference pipeline on ARM hardware, including latency and power measurements using TFLite Micro on the STM32H747XI Portenta H7 platform.

	\item We analyse deployment-oriented threshold calibration and show that mean recall of at least 98\% (Micro) and 99\% (Edge) is maintained after INT8 quantisation with negligible performance loss.
\end{itemize}

The results demonstrate that tropical bird audio detection can be achieved with models occupying less than 8\,kB of flash memory while maintaining the high recall required for practical field deployment. To support reproducible research, all models, training code, and evaluation protocols are released as open-source resources.

The remainder of this paper is organised as follows. Section~\ref{sec:background} reviews related work. Section~\ref{sec:problem} formulates the problem and defines the design constraints. Section~\ref{sec:dataset} describes the SEABAD dataset. Section~\ref{sec:architecture} presents the proposed architectures. Section~\ref{sec:experiments} details the training configuration, baselines, and ablation study. Section~\ref{sec:expsetup} describes the evaluation methodology, and Section~\ref{sec:results} presents the experimental results. Section~\ref{sec:deployment} evaluates embedded deployment, Section~\ref{sec:discussion} discusses implications and limitations, and Section~\ref{sec:conclusion} concludes the paper.

\section{Background and Related Work}
\label{sec:background}

\subsection{Edge-Based Bioacoustic Monitoring}

The shift from cloud-based processing to edge computing in passive acoustic monitoring is driven by the practical limitations of autonomous recording units (ARUs), which are typically battery- or solar-powered and often deployed without reliable network connectivity. Hardware has evolved from proprietary recorders to open platforms such as AudioMoth \citep{hill2019audiomoth}, while recent studies have demonstrated on-device inference on reprogrammed Cortex-M4 processors \citep{ciapponi2025enabling} and quantified the battery savings achievable through TinyML-triggered recording \citep{benhammadi2026battery}.

At the software level, TinyML enables neural inference on microcontroller-class hardware through frameworks such as TensorFlow Lite for Microcontrollers \citep{david2021tensorflow} and lightweight architectures including MobileNets \citep{howard2017mobilenets}. More recent Tiny Deep Learning approaches extend these ideas using quantisation, pruning, and neural architecture search to accommodate deeper networks within MCU resource limits \citep{somvanshi2025tiny}. These techniques have already been applied to tasks such as hornbill-call classification on Arduino-class hardware \citep{kongkahing2025hornbill} and the BioDCASE 2025 ESP32-S3 benchmark \citep{carmantini2025biodcase}. Nevertheless, many embedded bioacoustic systems still rely on models originally developed for multi-class classification, even when the deployment task is simply detecting the presence of bird vocalisations.

\subsection{Lightweight Bird Audio Detection}

Bird Audio Detection (BAD) emerged through community evaluation campaigns \citep{stowell2016bird,dcase2018task}, which established strong baselines using 10-second recordings sampled at 44.1\,kHz. Subsequent reviews have documented the rapid adoption of deep learning for this task \citep{stowell2022computational}. More recent resources such as BirdSet \citep{rauch2025birdset} provide standardised evaluation across geographically diverse datasets, while systems such as HawkEars \citep{huus2025hawkears} demonstrate high detection accuracy on long-duration recordings using server-class hardware.

The assumptions underlying these systems differ from those of embedded ARUs. Most bird vocalisations occur below 8\,kHz and can be detected from relatively short audio segments, making high sampling rates and long temporal windows unnecessary for binary presence detection. In addition, many published detectors target platforms such as Bela \citep{mcpherson2015bela}, which provide substantially greater computational resources than microcontroller-based devices such as AudioMoth. On embedded hardware, memory footprint, inference latency, and energy consumption become dominant design constraints.

Relatively few studies have addressed BAD under these constraints. TinyChirp \citep{huang2024tinychirp} proposed compact CNNs of approximately 25\,kB for Corn Bunting detection, BirdVoxDetect \citep{lostanlen2024birdvoxdetect} focuses on nocturnal flight calls, and the BioDCASE 2025 challenge benchmarks Yellowhammer detection on ESP32-S3 hardware \citep{carmantini2025biodcase}. Similar embedded implementations have been reported for hornbill-call detection \citep{kongkahing2025hornbill}, while quantisation techniques further reduce deployment costs \citep{solomes2020efficient}. Perez et al.\ provide a broader review of embedded acoustic monitoring systems \citep{perez2026wabad}. Most existing work, however, has focused on single-species detection or temperate datasets rather than species-rich tropical soundscapes.

\subsection{Embedded Architecture Design and Threshold Optimisation}

Several architectural techniques have become standard for reducing computational cost on embedded hardware. These include depthwise-separable convolutions \citep{majumdar2020matchboxnet,howard2017mobilenets}, replacing fully connected layers with global average pooling, and post-training INT8 quantisation \citep{jacob2018quantization}. Broader TinyML surveys also discuss pruning, neural architecture search, and hardware--software co-design for resource-constrained deployment \citep{somvanshi2025tiny}. Efficient execution on ARM Cortex-M processors is supported by inference libraries such as CMSIS-NN \citep{lai2018cmsisnn} and frameworks including MCUNet \citep{lin2020mcunet}. For imbalanced detection problems, focal loss \citep{lin2017focal} is frequently adopted because it reduces the influence of easily classified examples while placing greater emphasis on more difficult training instances.

Although these techniques are well established individually, comparatively little work has evaluated them together for tropical multi-species bird audio detection under strict microcontroller memory and latency budgets. This study investigates that design space through a systematic comparison of lightweight architectures and deployment-oriented optimisation strategies.

\section{Problem Formulation and Design Constraints}
\label{sec:problem}

\subsection{Task Definition}

Bird Audio Detection (BAD) is treated as a binary classification problem. Given a 3-second audio segment sampled at 16\,kHz, the objective is to determine whether it contains at least one bird vocalisation. The input to the network is a log-mel spectrogram, and the output is a probability $p \in [0,1]$. During inference, this probability is converted to a binary decision using a threshold $\tau$. A 3-second window captures the duration of most bird vocalisations (approximately 0.5--3\,s) while matching the buffering intervals commonly used by autonomous recording units.

\subsection{Design Constraints}

The primary deployment target is an ARM Cortex-M4-class microcontroller representative of open-source ARUs such as AudioMoth \citep{hill2019audiomoth}. The design is constrained by the limited memory and computational resources available on these platforms:

\begin{itemize}
	\item \textbf{Memory:} up to 4\,MB of on-chip Flash for program and model storage, and up to 256\,kB of SRAM for the tensor arena and runtime buffers.
	\item \textbf{Model size:} less than 10\,kB after INT8 quantisation for the Nano and Micro variants, leaving sufficient Flash for firmware, audio buffering, and optional downstream species-classification models.
	\item \textbf{Latency:} feature extraction and inference must complete well within the 3-second analysis window so that the compute pipeline never becomes the rate-limiting stage of duty-cycled recording. On a Cortex-M7-class platform (Portenta H7, our measurement vehicle) this target is comfortably met---the full mel-plus-inference pipeline runs in ${\sim}53$\,ms, a ${<}2\%$ compute duty cycle (Section~\ref{sec:deployment}). On the slower Cortex-M4F of an AudioMoth-class ARU the same pipeline is projected at ${\sim}480$\,ms (Section~\ref{sec:deployment}), still an order of magnitude inside the 3-second window and leaving the SD-write, not inference, as the dominant per-window cost.
\end{itemize}

Linux-based single-board computers, including the Raspberry Pi 4 and Portenta X8, are considered a secondary deployment platform. Their larger memory budgets permit models up to 50\,kB, while inference latency is constrained to below 2\,ms to support real-time streaming.

\subsection{Design Objectives}

The detector is designed to maximise recall within the available memory and latency budget. In long-term acoustic monitoring, missed detections permanently reduce survey coverage because missed vocalisations cannot be recovered after deployment. False positives are generally less costly, as they only increase the number of recordings passed to downstream processing.

Accordingly, the target recall---defined on the across-seed mean at the calibrated operating threshold---is at least 98\% for the microcontroller configuration (DrongoNet-Micro, AudioMoth-class Cortex-M4) and at least 99\% for the single-board-computer configuration (DrongoNet-Edge, Raspberry Pi/Linux ARUs). Concretely, at 98\% recall no more than 2 in every 100 bird-containing segments are missed; at 99\%, no more than 1 in 100. Both variants clear these floors on the mean (Micro 98.3\%, min-seed 97.9\%; Edge 99.0\%, min-seed 98.8\%; Section~\ref{sec:threshold}). The higher recall target for the SBC configuration reflects its less restrictive computational budget. Similar operating points have been adopted in bioacoustic monitoring applications where detection sensitivity is prioritised over reducing false positives \citep{stowell2022computational,lostanlen2024birdvoxdetect}.

AUC and F1-score are used to compare architectures during model development, whereas final model selection is based on recall at the calibrated decision threshold $\tau$. Any performance loss introduced by INT8 quantisation is required to remain below 0.5\% relative to the corresponding floating-point model.

\section{SEABAD Dataset}
\label{sec:dataset}

\subsection{Overview}

Experiments use the Southeast Asian Bird Audio Detection (SEABAD) dataset \citep{zabidi2026seabad}, a large-scale corpus of tropical bird recordings. Only the dataset characteristics relevant to model development are described here; details of the data collection and curation process are provided in Appendix~\ref{app:dataset}.

SEABAD contains 50,000 automatically curated 3-second audio clips sampled at 16\,kHz. The dataset is split into 40,000 training, 5,000 validation, and 5,000 test samples, with an equal number of positive and negative examples in each partition. A manual audit of 1,000 randomly selected clips estimated the label accuracy at 97.8\%\,$\pm$\,0.9\%.

\paragraph{Input Representation.}
Each audio clip is converted to a log-mel spectrogram \citep{davis1980comparison}. Using a hop length of 256 samples produces 184 time frames per recording. Three spectrogram configurations are evaluated: Nano ($n_\mathrm{fft}=512$, $n_\mathrm{mels}=16$), Micro ($n_\mathrm{fft}=1024$, $n_\mathrm{mels}=16$), and Edge ($n_\mathrm{fft}=1024$, $n_\mathrm{mels}=80$). These settings were selected based on the ablation study in Section~\ref{sec:mel_sweep}.

\subsection{Dataset Complexity and Capacity Requirements}
\label{sec:dataset_complexity}

SEABAD includes recordings from 1{,}677 tropical bird species and provides approximately 3.5 times as many training samples as the single-species TinyChirp Corn Bunting dataset. In contrast to single-species detection, the positive class spans a broad range of vocalisation types with diverse temporal and spectral characteristics. Since both datasets use 3-second recordings sampled at 16\,kHz, differences in detection performance can largely be attributed to task complexity rather than differences in the input representation. This makes SEABAD a suitable benchmark for evaluating how model capacity affects detection performance.

\section{DrongoNet Architecture Design}
\label{sec:architecture}

\subsection{Overview of the DrongoNet Framework}

\begin{figure}[!t]
	\centering
	\includegraphics[width=\textwidth]{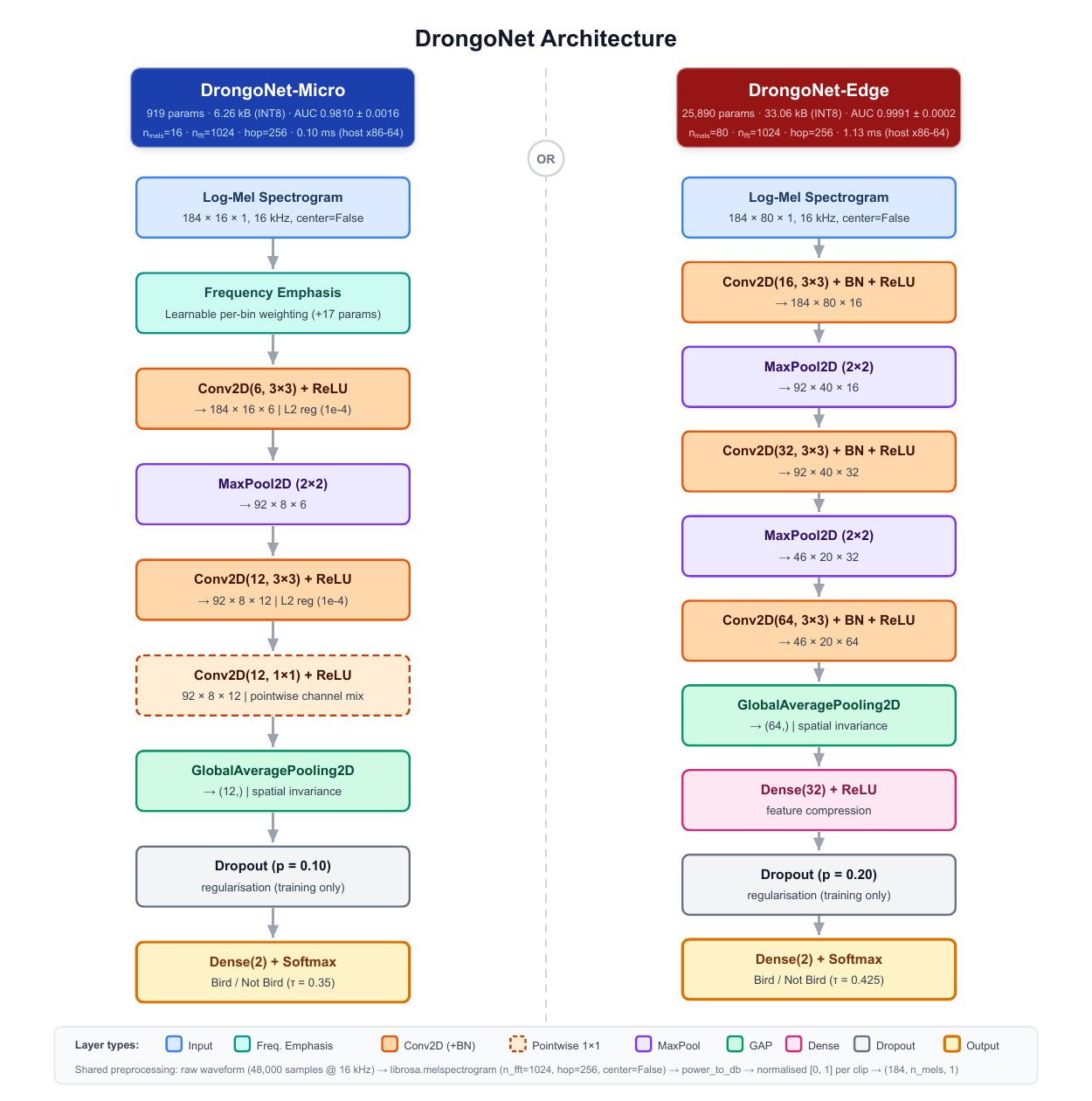}
	\caption{DrongoNet architecture, showing the two topologies DrongoNet-Micro (left) and DrongoNet-Edge (right); DrongoNet-Nano shares the Micro topology with $n_\mathrm{fft}{=}512$ (vs.\ 1024) and a trimmed filter count. Input is a 3-second 16\,kHz log-mel spectrogram with hop length 256. Micro applies a 17-parameter learnable frequency-emphasis layer (a quantisation-safe replacement for BN, Eq.~\ref{eq:freq_emph}) followed by two Conv2D stages and a pointwise 1$\times$1 channel-mix refinement; Edge stacks three Conv2D+BN blocks of increasing capacity (16$\to$32$\to$64 filters). Both branches collapse spatial features via global average pooling before a two-class dense softmax classifier, operated at $\tau{=}0.35$ (Micro) and $\tau{=}0.425$ (Edge; \S\ref{sec:threshold}).}
	\label{fig:architecture}
\end{figure}

DrongoNet comprises three lightweight CNN variants for bird presence detection on embedded devices. Figure~\ref{fig:architecture} illustrates the processing pipeline. A 3-second audio segment is converted into a log-mel spectrogram, optionally passed through a learnable frequency-emphasis layer, and processed by successive convolutional blocks. Global average pooling produces a compact feature representation, followed by a two-class softmax classifier.

The three variants (Nano, Micro, and Edge) share the same design philosophy but differ in spectrogram resolution and network capacity to accommodate different hardware platforms, from Cortex-M microcontrollers to Linux-based single-board computers. Their development was guided by a four-stage ablation study based on the TinyChirp CNN-Mel architecture \citep{huang2024tinychirp}, which evaluated the effect of individual architectural modifications.

\subsection{Design Principles}
\label{sec:design_principles}

The DrongoNet architecture is optimised for three critical constraints in embedded bioacoustic monitoring, derived from the acoustic properties of bird vocalisations: narrowband spectral structure (1--8\,kHz), short temporal duration (50--500\,ms), and repetitive temporal patterns that permit aggressive pooling without loss of onset-detection precision.

\paragraph{Temporal Resolution Preservation.}
Bird vocalisations are brief (50--500\,ms) and intermittent. The first convolutional layer uses a stride of 1 to preserve full temporal resolution; downsampling is deferred to the second layer after extracting fine-grained spectro-temporal features.

\paragraph{Parameter Efficiency.}
Global average pooling replaces parameter-intensive fully connected layers by averaging each feature map, substantially reducing model size. We also evaluated depthwise separable convolutions. Although they reduced parameter count, they produced lower mean AUC (0.9558 vs.\ 0.9824) and higher variance (±1.34 pp) than the standard-convolution GAP+focal base at $n_\mathrm{mels}=16$ (Table~\ref{tab:incremental_complexity}). Standard convolutions were therefore retained in the final architecture.

\paragraph{Edge Deployability and Quantisation Compatibility.}
The architecture uses only operations supported efficiently by TensorFlow Lite Micro (convolution, pooling, and ReLU), so INT8 post-training quantisation is straightforward. Spectral normalisation is handled in two ways across the family. DrongoNet-Edge retains batch normalisation, which folds cleanly into the preceding convolution weights at export time and improves generalisation with negligible quantisation loss. In the smaller Nano and Micro models, batch normalisation is replaced by a lightweight learnable frequency-emphasis layer inspired by adaptive audio front-ends \citep{zeghidour2021leaf} and channel-wise feature recalibration \citep{hu2018squeeze} but simplified for MCU deployment. Given a log-mel spectrogram $\mathbf{X} \in \mathbb{R}^{T \times F}$ (time $\times$ frequency bins), the layer computes
\begin{equation}
  \hat{\mathbf{X}} = \mathbf{X} \odot \sigma(\mathbf{w} \cdot s),
  \label{eq:freq_emph}
\end{equation}
where $\mathbf{w} \in \mathbb{R}^{F}$ are learnable per-bin weights (initialised to 1), $s \in \mathbb{R}$ is a learnable scale factor (initialised to 3), and $\sigma(\cdot)$ is the sigmoid. Bounding the weight map to $(0,1)$ lets the model suppress non-informative frequency bands during training rather than relying on the fixed mel filterbank boundaries. At $n_\mathrm{mels}{=}16$ this adds only $F{+}1{=}17$ trainable parameters, and inference reduces to integer multiplications.

\subsection{Architecture Details (Nano, Micro, Edge)}
\label{sec:arch_details}
Applying these design principles yields the three DrongoNet configurations summarised in Table~\ref{tab:model_variants}. Nano targets the smallest memory budget, Micro is the Cortex-M-class deployment target, and Edge scales up filter capacity for Linux SBCs. Runtime performance and deployment measurements are reported in Section~\ref{sec:results}.

\section{Experiments}
\label{sec:experiments}

We describe training setup, baseline selection, the four-phase ablation, and the resulting variant specifications.

\subsection{Training Configuration}

We trained with AdamW~\citep{loshchilov2019decoupled}: weight decay $10^{-4}$, initial learning rate $10^{-3}$ halved on validation-AUC plateau (patience 5, $\text{min\_lr}{=}10^{-5}$), batch size 32, up to 100 epochs with early stopping on validation AUC (patience 15, best-weight restoration). The loss is focal binary cross-entropy ($\gamma{=}2.0$, $\alpha{=}0.5$) for Nano and Micro; Edge uses standard categorical cross-entropy, since focal loss caused training instability at Edge's higher capacity and $n_\mathrm{mels}{=}80$ resolution. Online augmentation adds Gaussian noise ($\sigma{=}0.02$) to the log-mel input and applies a $\pm10$-frame circular shift along the time axis with probability 0.5. Each seed trained in 7--13\,min on a single GTX 1080 Ti (TensorFlow 2.15); the test set was held out throughout, and hyperparameters were tuned on the validation split.

INT8 quantisation uses the TFLite converter with a 500-sample calibration set drawn from validation. Default post-training quantisation (\texttt{tf.lite.Optimize.DEFAULT}) runs at the end of each training job; a separate full-INT8 pipeline (\texttt{OpsSet.TFLITE\_BUILTINS\_INT8} with int8 I/O) produces the CMSIS-NN-deployable models reported in Section~\ref{sec:results}. AUC degradation versus float32 stays below 0.01\% for Nano and Micro and below 0.12\% for Edge (0.11\% at the worst seed).

\subsection{Baseline Selection}
\label{sec:baseline}

The starting point is the TinyChirp CNN-Mel architecture~\citep{huang2024tinychirp}, a compact 2D CNN (25.6K parameters) reaching $99.91\%$ AUC on single-species Corn Bunting detection. Zero-shot on SEABAD, the pretrained weights collapse to chance ($55.65\% \pm 1.48\%$ AUC; the raw-waveform CNN-Time variant behaves identically at $43.36\% \pm 1.03\%$, Table~\ref{tab:tinychirp_seabad}), confirming that temperate single-species weights do not transfer to multi-species tropical detection. Retrained from scratch on SEABAD ($n_\mathrm{mels}{=}80$), the same architecture recovers to $0.9815 \pm 0.0008$ AUC---our fair equal-architecture baseline.

\begin{table*}[ht]
	\centering
	\caption{TinyChirp architectures on their original Corn Bunting task vs.\ SEABAD under zero-shot transfer (no retraining). The two AUC columns are evaluated on the Corn Bunting and SEABAD test sets respectively, using identical pre-trained weights; values are mean\,$\pm$\,std over three TinyChirp training seeds (42/100/786) with the SEABAD test split held fixed. Both architectures collapse to chance on SEABAD. Retrained from scratch ($n_\mathrm{mels}{=}80$), CNN-Mel recovers to $0.9815\pm0.0008$ AUC (Table~\ref{tab:baseline_comparison}).}
	\label{tab:tinychirp_seabad}
	\begin{tabular}{lccccc}
		\toprule
		\textbf{Architecture} & \textbf{Params} & \textbf{Peak act.\ (kB)$^\ast$} & \textbf{TinyChirp AUC (\%)} & \textbf{SEABAD AUC (\%)} & \textbf{$\Delta$ AUC} \\
		\midrule
		CNN-Mel & 25.6k & 55 & $99.91 \pm 0.10$ & $55.65 \pm 1.48$ & $-44.3$ \\
		CNN-Time & 826 & 188 & $99.16 \pm 0.09$ & $43.36 \pm 1.03$ & $-55.8$ \\
		\bottomrule
	\end{tabular}
	\smallskip\\
	{\footnotesize $^\ast$ Largest INT8 activation tensor---a lower bound on the TFLite-Micro tensor arena. CNN-Time's raw-waveform frontend convolves the 48{,}000-sample input directly, so despite its far smaller weight count its activation footprint is the larger of the two; see text.}
\end{table*}

We build on CNN-Mel rather than CNN-Time: although CNN-Time has 31$\times$ fewer weights, its raw-waveform frontend produces a $\sim$188\,kB first-layer activation that is $6\times$ AudioMoth's 32\,kB SRAM and difficult to reduce within typical MCU memory budgets. While streaming inference and quantisation-to-int8 approaches can reduce activation footprints, they are beyond the scope of this paper; we therefore focus on mel-frontend architectures, which are more readily deployable on resource-constrained hardware. The mel frontend's $184{\times}80$ activation is reducible to a 23\,kB arena at $n_\mathrm{mels}{=}16$, and replacing CNN-Mel's parameter-heavy \texttt{Reshape}+\texttt{Dense} classifier with GAP (Section~\ref{sec:gap_focal}) further shrinks the model by an order of magnitude at no accuracy cost.

Table~\ref{tab:baseline_comparison} places DrongoNet in the broader landscape: Micro achieves 0.9810 AUC in 6.26\,kB---over $1{,}200\times$ fewer parameters than MobileNetV3-Small---while Edge matches ResNet-50's AUC in 33.06\,kB ($730\times$ smaller).

\begin{table*}[ht]
	\centering
	\caption{Baseline comparison: DrongoNet family vs.\ standard architectures fine-tuned on SEABAD. AUC: mean\,$\pm$\,std across five random seeds (42, 100, 786, 7, 1234) for DrongoNet variants; standard architectures across three seeds (42, 100, 786). Standard model sizes are approximate INT8 estimates ($\approx$1 byte/param).}
	\label{tab:baseline_comparison}
	\small
	\begin{tabular}{lcccccc}
		\toprule
		\textbf{Model} & \textbf{Params} & \textbf{Size (KB)} & \textbf{Input} & \textbf{AUC} & \textbf{Platform} & \textbf{Fits AudioMoth} \\
		\midrule
		VGG-16$^\dagger$          & 14.9M  & $\sim$14,900 & $224{\times}224$ & $0.9995 \pm 0.0001$ & GPU/Server & $\times$ \\
		ResNet-50$^\dagger$        & 24.2M  & $\sim$24,200 & $224{\times}224$ & $0.9992 \pm 0.0003$ & GPU/Server & $\times$ \\
		EfficientNet-B0$^\dagger$  & 4.4M   & $\sim$4,400  & $224{\times}224$ & $0.9991 \pm 0.0004$ & SBC/Mobile & $\times$ \\
		MobileNetV3-Small$^\dagger$& 1.1M   & $\sim$1,100  & $224{\times}224$ & $0.9985 \pm 0.0002$ & SBC/Mobile & $\times$ \\
		MicroNet-KWS-S$^\ddagger$  & 62.6K    & 98.09    & $184{\times}80$  & $0.9986 \pm 0.0004$ & MCU (M4 class) & $\times$ \\
		TinyChirp CNN-Mel$^\S$     & 25.6K    & 28.61    & $184{\times}80$  & $0.9815 \pm 0.0008$ & MCU (M4 class) & $\times$ \\
		\midrule
		\textbf{DrongoNet-Nano}  & \textbf{763}    & \textbf{5.09}  & $184{\times}16$ & $\mathbf{0.9727 \pm 0.0016}$ & \textbf{MCU (M4 class)} & \checkmark \\
		\textbf{DrongoNet-Micro} & \textbf{919}    & \textbf{6.26}  & $184{\times}16$ & $\mathbf{0.9810 \pm 0.0016}$ & \textbf{MCU (M4 class)} & \checkmark \\
		\textbf{DrongoNet-Edge}  & \textbf{25,890} & \textbf{33.06} & $184{\times}80$ & $\mathbf{0.9991 \pm 0.0002}$ & \textbf{Linux SBC} & $\times$ \\
		\bottomrule
	\end{tabular}
	\smallskip\\
	{\footnotesize $^\dagger$ImageNet-pretrained, fine-tuned on SEABAD at 224$\times$224 input (3 seeds); the resolution mismatch to DrongoNet makes the AUC comparison approximate. $^\ddagger$MicroNet-KWS-S and $^\S$TinyChirp retrained from scratch on SEABAD; both, and the MCUNet reference, are discussed in the text. Sizes for standard models are INT8 estimates ($\approx$1\,byte/param). ``Fits AudioMoth'' requires flash $\leq$256\,kB \emph{and} SRAM $\leq$32\,kB.}
\end{table*}

The two dedicated MCU baselines are reproduced under SEABAD's requirements rather than cited from their source papers. MicroNet-KWS-S~\citep{banbury2021micronets} is faithfully rebuilt from its Table~5 (Conv2D stem plus five depthwise-separable blocks), substituting global average pooling for the fixed-shape AvgPool and cross-entropy loss (the source paper carries no bird-audio recipe), and retrained on SEABAD over three seeds. Although its 98.09\,kB flash footprint fits comfortably, at SEABAD's $184{\times}80$ input its peak INT8 activation reaches 1207.5\,kB---about $38\times$ AudioMoth's 32\,kB SRAM---so it cannot deploy on the target hardware; its published configuration already assumes an STM32F446RE with $4\times$ AudioMoth's SRAM budget.

MCUNet~\citep{lin2020mcunet}'s smallest published variant likewise targets 256\,kB SRAM ($8\times$ AudioMoth's budget), despite sharing the Cortex-M4 core family, and is not retrained here. TinyChirp CNN-Mel, retrained on SEABAD at $n_\mathrm{mels}{=}80$, has a 55\,kB peak activation that also exceeds the 32\,kB ceiling (Table~\ref{tab:tinychirp_seabad}). Only DrongoNet-Nano and -Micro, whose arenas measure ${\approx}23$\,kB (Table~\ref{tab:inference_latency}), clear both the flash and SRAM bars---hence the ``Fits AudioMoth'' column.


\subsection{Ablation Overview}

\begin{figure}[!t]
  \centering
  \includegraphics[width=\textwidth]{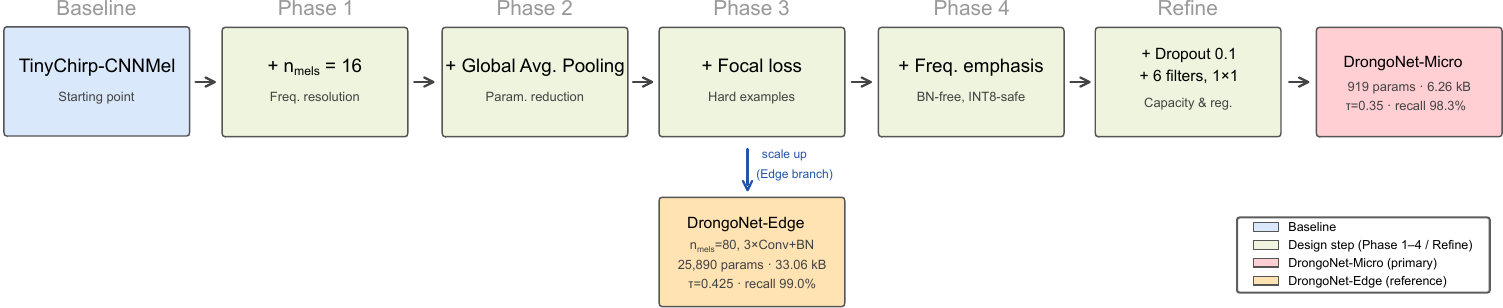}
  \caption{Ablation pipeline from TinyChirp-CNNMel to DrongoNet-Micro (left to right). Each box represents one cumulative design change; phase labels (Phase 1--4, then Refine) are shown above. The Edge branch forks at Phase~1, retaining $n_\mathrm{mels}{=}80$ and scaling up filter capacity with three Conv2D+BN blocks while skipping Phase~4. Strided and depthwise-separable convolutions were evaluated at the Refine step but rejected due to AUC degradation at the $n_\mathrm{mels}{=}16$ scale.}
  \label{fig:ablation_pipeline}
\end{figure}

Starting from the TinyChirp-CNNMel baseline, the Micro branch is derived through four cumulative phases plus a final refinement step, as shown in Figure~\ref{fig:ablation_pipeline}. The Edge branch forks at Phase~1 by retaining the high-resolution $n_\mathrm{mels}{=}80$ setting instead of dropping to 16.

\begin{enumerate}
    \item \textbf{Phase 1 --- Frequency resolution} (Table~\ref{tab:mel_sweep}): a mel-bin sweep over both FFT sizes shows AUC peaks at $n_\mathrm{mels}{=}80$ / $n_\mathrm{fft}{=}1024$ (0.9957), fixing the Edge configuration. Dropping to $n_\mathrm{mels}{=}16$ trades ${\sim}1$\,pp of AUC for an order-of-magnitude smaller activation footprint, defining the MCU-scale Micro branch that the remaining phases refine.
    \item \textbf{Phase 2 --- Global Average Pooling} (Table~\ref{tab:gap_ablation}, first two rows): replacing the parameter-heavy Flatten+Dense classifier with GAP cuts model size by $44\%$ but costs 1.52\,pp of AUC at $n_\mathrm{mels}{=}16$. The loss is recovered in Phase~3.
    \item \textbf{Phase 3 --- Focal loss} (Table~\ref{tab:gap_ablation}, last row): swapping cross-entropy for focal loss ($\gamma{=}2.0$) on top of GAP recovers the Phase-2 loss and exceeds the Flatten+CE baseline, reaching $98.24\% \pm 0.12\%$ AUC at 5.37\,kB.
    \item \textbf{Phase 4 --- Frequency emphasis} (Table~\ref{tab:fe_ablation}): adding the 17-parameter learnable spectral-weighting layer leaves AUC essentially unchanged ($-0.07 \pm 0.10$\,pp). The motivation is not accuracy but quantisation: it provides a BN-free, INT8-safe replacement for batch normalisation, which degrades under INT8 at this scale.
    \item \textbf{Refine} (Table~\ref{tab:incremental_complexity}): the final Micro adds dropout 0.1, trims to 6 filters with a 1$\times$1 refinement stage, and is checked against strided and depthwise-separable alternatives (both rejected: strided costs 0.76\,pp, depthwise-separable degrades to $95.58\% \pm 1.34\%$ with high seed variance).
\end{enumerate}

The deployed variants combine these phases: Edge uses Conv2D+BN blocks + GAP + standard cross-entropy at $n_\mathrm{mels}{=}80$ (forking at Phase~1, no Phase~4; focal loss was tested but caused training instability at this capacity, so plain cross-entropy was retained instead); Micro uses Phases 1--4 plus Refine---Conv2D + GAP + focal loss + frequency emphasis at $n_\mathrm{mels}{=}16$, no BN. Both are validated across five seeds.

\subsection{Frequency Resolution Optimisation}
\label{sec:mel_sweep}

Table~\ref{tab:mel_sweep} sweeps mel bins across both FFT sizes. AUC rises with bin count and peaks at $n_\mathrm{mels}{=}80$ for both 1024-FFT (0.9957) and 512-FFT (0.9804). From this sweep we fix Edge at $n_\mathrm{mels}{=}80$ with 1024-FFT, Micro at $n_\mathrm{mels}{=}16$ with 1024-FFT, and Nano at $n_\mathrm{mels}{=}16$ with 512-FFT---trading ${\approx}0.8$\,pp of deployed AUC (Table~\ref{tab:model_variants}) for a ${\sim}40\%$ reduction in mel preprocessing latency relative to Micro.

\begin{table*}[t]
    \centering
    \caption{Mel bin sweep (mean\,$\pm$\,std across seeds 42, 100, 786; TFLite INT8). GAP + focal loss + frequency emphasis architecture throughout---this is the single-block Micro-family architecture, \emph{not} the 3-block Edge network; the $n_\mathrm{mels}{=}80$ latency here (0.44\,ms) therefore differs from the deployed DrongoNet-Edge (1.13\,ms, Table~\ref{tab:model_variants}), which adds two further Conv2D+BN blocks.}
    \label{tab:mel_sweep}
    \smallskip
    \begin{tabular}{@{}cccccc@{}}
        \toprule
        \textbf{$n_\mathrm{fft}$} & \textbf{$n_\mathrm{mels}$} & \textbf{FP32 AUC} & \textbf{INT8 AUC} & \textbf{Accuracy} & \textbf{Latency (ms)} \\
        \midrule
        \multirow{5}{*}{1024}
        & \textbf{80} & $\mathbf{0.9957\pm0.0004}$ & $\mathbf{0.9956\pm0.0005}$ & \textbf{97.25\%} & \textbf{0.44} \\
        & 64 & $0.9890\pm0.0022$ & $0.9888\pm0.0023$ & 95.09\% & 0.35 \\
        & 48 & $0.9854\pm0.0012$ & $0.9854\pm0.0013$ & 94.07\% & 0.27 \\
        & 32 & $0.9844\pm0.0027$ & $0.9843\pm0.0027$ & 93.99\% & 0.18 \\
        & \textbf{16} & $\mathbf{0.9848\pm0.0011}$ & $\mathbf{0.9846\pm0.0010}$ & 94.16\% & \textbf{0.09} \\
        \midrule
        \multirow{5}{*}{512}
        & 80 & $0.9804\pm0.0025$ & $0.9800\pm0.0027$ & 93.17\% & 0.38 \\
        & 64 & $0.9802\pm0.0005$ & $0.9801\pm0.0007$ & 93.23\% & 0.31 \\
        & 48 & $0.9776\pm0.0009$ & $0.9774\pm0.0008$ & 92.87\% & 0.23 \\
        & 32 & $0.9721\pm0.0035$ & $0.9717\pm0.0034$ & 91.96\% & 0.15 \\
        & \textbf{16} & $\mathbf{0.9595\pm0.0075}$ & $\mathbf{0.9593\pm0.0074}$ & 89.90\% & \textbf{0.07} \\
        \bottomrule
    \end{tabular}
\end{table*}

\subsection{Global Average Pooling and Focal Loss}
\label{sec:gap_focal}

With Edge's resolution fixed by the sweep above, the remaining phases target the MCU-scale Micro branch at $n_\mathrm{mels}{=}16$. Following MobileNets~\citep{howard2017mobilenets}, we replace the parameter-heavy Flatten classifier with Global Average Pooling (GAP)~\citep{lin2013network}. Table~\ref{tab:gap_ablation} shows that GAP alone costs 1.52\,pp of AUC, but pairing it with focal loss~\citep{lin2017focal} recovers that loss and exceeds the Flatten+CE baseline.

\begin{table*}[ht]
	\centering
	\caption{GAP and focal loss ablation ($n_\mathrm{mels}{=}16$). AUC: mean\,$\pm$\,std across seeds 42, 100, 786.}
	\label{tab:gap_ablation}
	\begin{tabular}{llccc}
		\toprule
		\textbf{Model} & \textbf{Configuration} & \textbf{AUC (\%, mean$\pm$std)} & \textbf{$\Delta$} & \textbf{Size (KB)} \\
		\midrule
		Start & Flatten + CE & $97.06\pm0.11$ & --- & 7.30 \\
		GAP & GAP + CE & $95.54\pm0.12$ & $-1.52$ & 4.10 \\
		GAP+Focal & GAP + Focal & $\mathbf{98.24\pm0.12}$ & $\mathbf{+1.18}$ & 5.37 \\
		\bottomrule
	\end{tabular}
\end{table*}

Focal loss redirects gradient from easy examples toward borderline cases (faint calls, call-like noise), compensating for the spatial information lost to global pooling and improving recall at low operating thresholds (Section~\ref{sec:threshold}).

\subsection{Frequency Emphasis}
\label{sec:fe_ablation}
Table~\ref{tab:fe_ablation} compares GAP+Focal with and without the learnable spectral weighting layer (Eq.~\ref{eq:freq_emph}). The AUC difference is negligible ($-0.07 \pm 0.10$\,pp), so the layer is not motivated by accuracy. Its role in the shipped Micro is to \emph{replace} batch normalisation, which degrades under INT8 at this scale, with a quantisation-friendly alternative that adds only 17 parameters.

\begin{table}[ht]
	\centering
	\caption{FrequencyEmphasis ablation ($n_\mathrm{mels}{=}16$, $n_\mathrm{fft}{=}1024$). AUC (\%): mean\,$\pm$\,std across seeds 42, 100, 786. The layer does not improve AUC; its role is to replace BN as a quantisation-safe spectral normalisation.}
	\label{tab:fe_ablation}
	\begin{tabular}{lccc}
		\toprule
		\textbf{Variant} & \textbf{AUC (mean$\pm$std)} & \textbf{Size (KB)} & \textbf{$\Delta$ pp} \\
		\midrule
		GAP + Focal (no FE)   & $98.24 \pm 0.12$ & 5.37 & --- \\
		GAP + Focal + FE      & $98.17 \pm 0.05$ & 5.85 & $-0.07$ \\
		\bottomrule
	\end{tabular}
\end{table}

\subsection{Refine}
\label{sec:refine}

The Refine step in Figure~\ref{fig:ablation_pipeline} both validates the alternatives that were considered and rejected, and fixes the final Micro configuration. Table~\ref{tab:incremental_complexity} evaluates strided and depthwise-separable variants at $n_\mathrm{mels}{=}16$: strided convolutions cost 0.76\,pp of AUC, and removing the 1$\times$1 refinement costs a further 0.64\,pp. Depthwise-separable convolutions with BN yield the worst result, $95.58\% \pm 1.34\%$, with seed variance an order of magnitude larger than the other rows. This confirms that depthwise factorisation is counter-productive at the $n_\mathrm{mels}{=}16$ spatial scale. We therefore retain GAP+Focal+FE (Phases~1--4) as the architectural template for Micro.

The shipped Micro then adds dropout 0.1 and trims to the 6-filter, 1$\times$1-refined configuration that fits the $<$8\,kB INT8 budget, reaching $0.9810 \pm 0.0016$ AUC. The ${\approx}0.14$\,pp gap below the Phase-3 ablation peak ($98.24\%$) is the price of quantisation stability ($<$0.01\% INT8 degradation) inside the deployment footprint.

\begin{table}[ht]
	\centering
	\caption{Incremental architecture evolution ($n_\mathrm{mels}{=}16$). AUC: mean\,$\pm$\,std across seeds 42, 100, 786.}
	\label{tab:incremental_complexity}
	\begin{tabular}{llcccc}
		\toprule
		\textbf{Model} & \textbf{Change} & \textbf{AUC (\%, mean$\pm$std)} & \textbf{Size (KB)} & \textbf{Latency (ms)} \\
		\midrule
		Base & GAP + Focal & $98.24\pm0.12$ & 5.37 & 0.08 \\
		Strided & + Strided Conv & $97.48\pm0.26$ & 6.51 & 0.05 \\
		No 1$\times$1 & Remove 1$\times$1 & $96.84\pm0.28$ & 5.23 & 0.04 \\
		Sep+BN & Depthwise Sep+BN & $95.58\pm1.34$ & 7.60 & 0.09 \\
		\bottomrule
	\end{tabular}
\end{table}

\subsection{Accuracy Ceiling Estimation}

To test whether additional capacity could push Edge higher, we trained two larger CNNs across three seeds: a four-block network ($[16,32,64,128]$ filters, 106\,K params) reaches $0.9979 \pm 0.0011$ AUC, and a wider variant ($[32,64,128,256]$, 423\,K params) reaches $0.9984 \pm 0.0003$---neither exceeding DrongoNet-Edge's $0.9991 \pm 0.0002$. AUC plateaus near $99.8$--$99.9\%$ across a 16$\times$ capacity range, suggesting the residual gap is dataset-limited rather than capacity-limited. Adding WrenNet's semi-learnable frequency map~\citep{ciapponi2025enabling} to Edge as a learned frequency front-end shows the same pattern: $0.9983 \pm 0.0003$ AUC (INT8, three seeds) versus the control's $0.9983 \pm 0.0001$---within noise, reinforcing that the ceiling holds regardless of where the added capacity goes.

\subsection{Final Variant Specifications}

Combining the choices from each phase yields the three DrongoNet variants in Table~\ref{tab:model_variants}, shown alongside the retrained TinyChirp baseline for direct comparison.

\begin{table}[ht]
	\centering
	\caption{DrongoNet deployment model variants (INT8-quantised). Three variants span an accuracy--size trade-off from ultra-low-flash (Nano) to near-ceiling (Edge). AUC: mean\,$\pm$\,std across five seeds 42, 100, 786, 7, 1234. Latency: TFLite INT8, CPU (host x86-64), 1000-call mean, excluding mel preprocessing; on-device inference (Portenta H7 bare-metal) is 21\,ms (Micro) and 398\,ms (Edge), see Table~\ref{tab:inference_latency}.}
	\label{tab:model_variants}
	\small
	\begin{tabular}{lcccccc}
		\toprule
		\textbf{Variant} & \textbf{$n_\mathrm{fft}$} & \textbf{$n_\mathrm{mels}$} & \textbf{AUC} & \textbf{Size (KB)} & \textbf{Latency (ms)} & \textbf{Target Hardware} \\
		\midrule
		TinyChirp$^\dagger$ (baseline) & 1024 & 80 & $0.9815 \pm 0.0008$ & 28.61 & 0.34 & Reference \\
		DrongoNet-Nano  & 512  & 16 & $0.9727 \pm 0.0016$ & 5.09  & 0.09 & MCU, M4 class (ultra-low flash) \\
		DrongoNet-Micro & 1024 & 16 & $0.9810 \pm 0.0016$ & 6.26  & 0.10 & MCU, M4 class \\
		DrongoNet-Edge  & 1024 & 80 & $0.9991 \pm 0.0002$ & 33.06 & 1.13 & Linux SBC / RPi \\
		\bottomrule
	\end{tabular}
	\smallskip\\
	{\footnotesize $^\dagger$TinyChirp CNN-Mel~\citep{huang2024tinychirp} retrained from scratch on SEABAD in its native configuration ($n_\mathrm{mels}{=}80$, $n_\mathrm{fft}{=}1024$, 25{,}558 params); zero-shot, the original weights fall to chance on SEABAD (Table~\ref{tab:tinychirp_seabad}).
	Nano and Micro sizes are from full INT8 TFLite models (\texttt{OpsSet.TFLITE\_BUILTINS\_INT8}, input/output int8) suitable for MCU deployment via TensorFlow Lite Micro and CMSIS-NN. Edge uses full INT8 in all reported results. AUC values are from the corresponding float32 training runs; separately converted full INT8 models (same architecture and seeds) show ${<}0.01\%$ AUC degradation for Nano/Micro and ${<}0.12\%$ for Edge.}
\end{table}

The three variants partition the deployment space rather than competing on a single axis. Nano targets severely flash-constrained MCUs where every kilobyte of model size counts. Micro is the recommended Cortex-M4-class detector for duty-cycled AudioMoth-style ARUs, trading 1.8\,pp of AUC for a $12.6\times$ smaller tensor arena than Edge (23\,kB vs.\ 290\,kB). Edge targets Linux SBCs where the 290\,kB arena is unconstrained and the higher AUC justifies the additional compute.

\section{Evaluation Methodology}
\label{sec:expsetup}

\begin{figure}[!b]
	\centering
	\includegraphics[width=\textwidth]{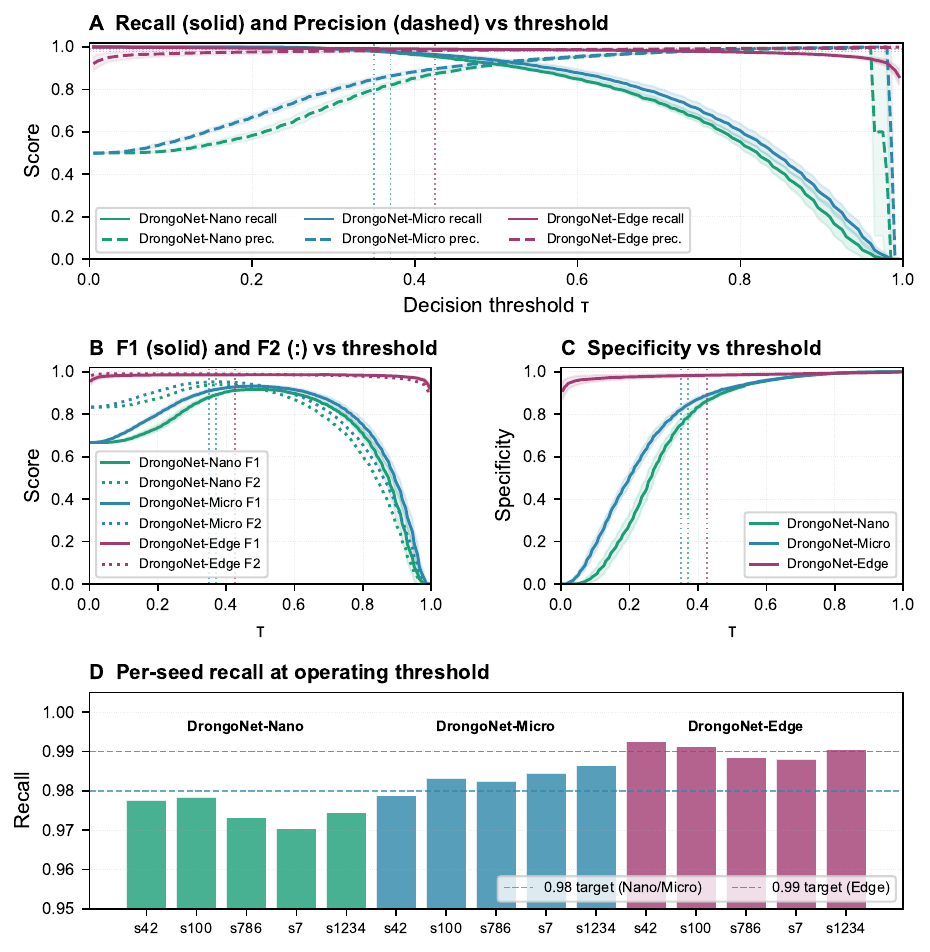}
	\caption{%
		\textbf{Threshold-dependent performance for DrongoNet-Micro and DrongoNet-Edge.}
		Shaded bands: $\pm$1\,std across seeds 42, 100, 786, 7, 1234. Dashed verticals mark
		$\tau{=}0.35$ (Micro) and $\tau{=}0.425$ (Edge, uniform across all five seeds).
		\textbf{(A)}~Recall and precision vs.~$\tau$.
		\textbf{(B)}~F$_1$ and F$_2$ scores.
		\textbf{(C)}~Sensitivity--specificity trade-off.
		\textbf{(D)}~Per-seed recall at the operating threshold.
	}
	\label{fig:threshold_analysis}
\end{figure}

\subsection{Metrics}

The DrongoNet variants are evaluated on the held-out SEABAD test set (5{,}000 clips, balanced 1:1) across five random seeds (42, 100, 786, 7, 1234); baseline architectures use three seeds (42, 100, 786). We report four quantities throughout:

\textbf{AUC (primary):} area under the ROC curve, mean\,$\pm$\,std across seeds.

\textbf{Recall at the operating threshold:} the fraction of bird vocalisations correctly identified at a per-variant $\tau$ tuned by the procedure of Section~\ref{sec:math}. For Nano and Micro, focal loss compresses output probabilities downward, pushing the operating $\tau$ below the default $\tau{=}0.5$; for Edge (standard cross-entropy), $\tau$ is likewise below 0.5, driven directly by the recall-floor selection rule of Section~\ref{sec:math} rather than loss-induced compression.

\textbf{INT8 quantisation degradation:} AUC difference between the float32 training run and the post-training INT8 TFLite model.

\textbf{Inference latency:} mean over 1,000 calls using TFLite INT8 on the host CPU, excluding mel preprocessing. On-device bare-metal measurements (Portenta H7) are in Table~\ref{tab:inference_latency}; mel preprocessing latency in Table~\ref{tab:melcalc}.

\subsection{Multi-Seed Evaluation Protocol}

Each DrongoNet variant is trained at five seeds (42, 100, 786, 7, 1234) and each baseline architecture at three (42, 100, 786), using identical hyperparameters and the same 40{,}000/5{,}000/5{,}000 train/val/test split throughout. The seed controls only weight initialisation and the random draws inside the data-augmentation pipeline, so the reported $\pm$std quantifies sensitivity to initialisation rather than to dataset variation.

\subsection{Threshold Selection}
\label{sec:math}


Well-tuned biodiversity monitoring pipelines routinely reach ${\geq}97\%$ recall (often 0.99) \citep{fleure2025impact}, making such floors appropriate for a gatekeeper stage. To pick a per-variant operating threshold we swept $\tau \in [0.05, 0.95]$ in 0.05 steps (refined to 0.025 near the recall floor) and selected the \emph{largest} $\tau$ at which the \emph{mean} recall across seeds still meets the recall floor---the largest $\tau$ maximises precision (and thus storage savings) while honouring the design target:
\begin{equation*}
  \tau_{\text{op}} = \max\!\Big\{\tau : \operatorname*{mean}_{s \in \{42,100,786,7,1234\}} \text{Recall}_s(\tau) \geq r_{\min}\Big\},
\end{equation*}
with $r_{\min}{=}0.98$ for Micro and $r_{\min}{=}0.99$ for Edge. This gives $\tau{=}0.35$ for Micro (mean recall 98.30\%, per-seed 97.88--98.64\%) and $\tau{=}0.425$ for Edge (mean recall 99.01\%, per-seed 98.80--99.24\%), both clearing their design targets on the mean. Practitioners can raise $\tau$ at inference time to trade recall for precision without retraining.

\section{Results and Analysis}
\label{sec:results}

\begin{figure}[!b]
  \centering
  \includegraphics[width=\textwidth]{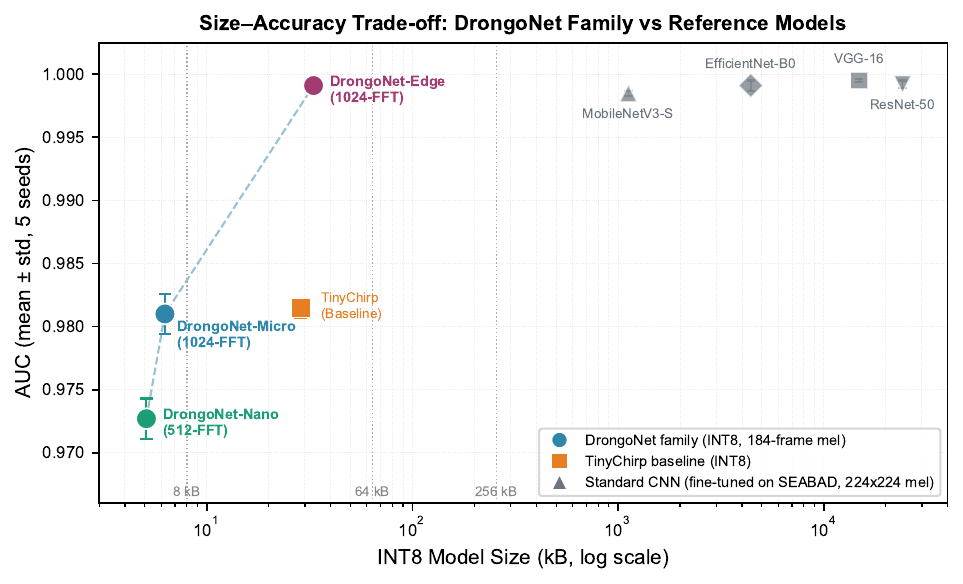}
  \caption{Size--accuracy trade-off on SEABAD test set. DrongoNet-Nano (5.09\,kB), Micro (6.26\,kB), and Edge (33.06\,kB) are INT8-quantised; error bars show $\pm$1\,std across five seeds. The retrained TinyChirp CNN-Mel baseline (28.6\,kB, $0.9815$ AUC) is shown for reference: at essentially the same footprint, DrongoNet-Edge reaches $0.9991$, while Nano/Micro retain comparable AUC at ${\sim}5\times$ smaller size. Standard CNNs (VGG-16, ResNet-50, EfficientNet-B0, MobileNetV3-S) were fine-tuned on SEABAD with 224$\times$224 mel input and ImageNet-pretrained weights; direct AUC comparison is approximate due to different input resolution. Standard model sizes are approximate INT8 estimates ($\approx$1\,byte/param). Vertical lines mark common MCU Flash budgets.}
  \label{fig:pareto}
\end{figure}

We evaluate all three variants on the held-out SEABAD test set (5{,}000 clips, five seeds) and present the results in four passes. Figure~\ref{fig:pareto} positions DrongoNet on the size--accuracy curve against the retrained TinyChirp baseline and four standard CNNs. Table~\ref{tab:perf_table} then breaks Micro and Edge out across the full $\tau$ sweep, anchoring the operating points used in the rest of the section. Section~\ref{sec:threshold} commits to those points; the remaining subsections analyse what they buy in discrimination, probability calibration, and deployment economics.

\begin{table*}[!t]
	\centering
	\caption{Performance metrics across decision thresholds for DrongoNet-Micro (operating $\tau{=}0.35$,
		bold) and DrongoNet-Edge. Values are mean\,$\pm$\,std across five random seeds (42, 100, 786, 7, 1234)
		on the held-out SEABAD test set (5{,}000 clips), full-INT8 models. $^\dagger$Marks the Edge operating
		threshold ($\tau{=}0.425$, the largest $\tau$ clearing the ${\geq}99\%$ mean-recall floor).}
	\label{tab:perf_table}
	\smallskip\footnotesize

	\begin{tabular}{@{}cccccc@{}}
		\toprule
		\multicolumn{6}{c}{\textbf{DrongoNet-Micro}} \\
		\cmidrule(lr){1-6}
		$\tau$ & Recall & Precision & Specificity & F$_1$ & F$_2$ \\
		\midrule
		0.05 & $1.000\pm0.000$ & $0.511\pm0.003$ & $0.043\pm0.012$ & $0.676\pm0.003$ & $0.839\pm0.002$ \\
		0.10 & $1.000\pm0.000$ & $0.551\pm0.009$ & $0.184\pm0.031$ & $0.710\pm0.008$ & $0.860\pm0.005$ \\
		0.20 & $0.999\pm0.001$ & $0.669\pm0.013$ & $0.506\pm0.028$ & $0.801\pm0.009$ & $0.909\pm0.004$ \\
		0.30 & $0.992\pm0.001$ & $0.797\pm0.014$ & $0.747\pm0.021$ & $0.884\pm0.008$ & $0.945\pm0.004$ \\
		\textbf{0.35} & $\mathbf{0.983\pm0.003}$ & $\mathbf{0.847\pm0.012}$ & $\mathbf{0.823\pm0.016}$ & $\mathbf{0.910\pm0.007}$ & $\mathbf{0.953\pm0.003}$ \\
		0.40 & $0.971\pm0.003$ & $0.883\pm0.009$ & $0.871\pm0.011$ & $0.925\pm0.005$ & $0.952\pm0.003$ \\
		0.50 & $0.938\pm0.009$ & $0.927\pm0.004$ & $0.926\pm0.005$ & $0.932\pm0.004$ & $0.936\pm0.007$ \\
		\bottomrule
	\end{tabular}

	\vspace{1em}

	\begin{tabular}{@{}cccccc@{}}
		\toprule
		\multicolumn{6}{c}{\textbf{DrongoNet-Edge}} \\
		\cmidrule(lr){1-6}
		$\tau$ & Recall & Precision & Specificity & F$_1$ & F$_2$ \\
		\midrule
		0.05 & $0.998\pm0.001$ & $0.961\pm0.019$ & $0.960\pm0.021$ & $0.979\pm0.010$ & $0.991\pm0.004$ \\
		0.10 & $0.996\pm0.001$ & $0.970\pm0.014$ & $0.969\pm0.015$ & $0.983\pm0.007$ & $0.991\pm0.002$ \\
		0.20 & $0.994\pm0.002$ & $0.976\pm0.010$ & $0.975\pm0.010$ & $0.985\pm0.004$ & $0.991\pm0.001$ \\
		0.30 & $0.993\pm0.001$ & $0.980\pm0.008$ & $0.980\pm0.009$ & $0.986\pm0.004$ & $0.990\pm0.001$ \\
		0.35 & $0.992\pm0.001$ & $0.982\pm0.007$ & $0.981\pm0.007$ & $0.987\pm0.003$ & $0.990\pm0.001$ \\
		0.40 & $0.991\pm0.002$ & $0.983\pm0.007$ & $0.983\pm0.007$ & $0.987\pm0.003$ & $0.989\pm0.002$ \\
		$0.425^\dagger$ & $0.990\pm0.002$ & $0.984\pm0.006$ & $0.984\pm0.006$ & $0.987\pm0.002$ & $0.989\pm0.001$ \\
		0.45 & $0.989\pm0.002$ & $0.984\pm0.006$ & $0.984\pm0.006$ & $0.987\pm0.002$ & $0.988\pm0.001$ \\
		0.50 & $0.989\pm0.002$ & $0.985\pm0.005$ & $0.985\pm0.006$ & $0.987\pm0.002$ & $0.988\pm0.001$ \\
		0.55 & $0.987\pm0.002$ & $0.987\pm0.005$ & $0.987\pm0.005$ & $0.987\pm0.002$ & $0.987\pm0.001$ \\
		0.60 & $0.986\pm0.002$ & $0.988\pm0.004$ & $0.988\pm0.004$ & $0.987\pm0.002$ & $0.986\pm0.001$ \\
		\bottomrule
	\end{tabular}
\end{table*}

The bold Micro row of Table~\ref{tab:perf_table} echoes the operating-point values in Table~\ref{tab:threshold_comparison}; the daggered Edge row marks $\tau{=}0.425$, the largest $\tau$ at which mean recall clears the ${\geq}99.0\%$ floor. The F$_2$ column (which weights recall twice as heavily as precision) is informative for the gatekeeper trade-off. Micro's F$_2$ peaks at the operating $\tau{=}0.35$ (0.953), so the recall-floor choice coincides with the F$_2$-optimal point, consistent with our gatekeeper-not-classifier framing. For Edge, near-perfect discrimination across the full sweep means the operating point clears the ${\geq}99\%$ recall target while retaining ${>}98\%$ precision.

\subsection{Threshold Selection and Operating Points}
\label{sec:threshold}

For Nano and Micro, focal loss compresses the output probability scale downward~\citep{lin2017focal}; for Edge (standard cross-entropy) no such compression applies. In both cases the default $\tau{=}0.5$ is not the optimal operating point, so a per-variant choice via the rule of Section~\ref{sec:math} is required.

\paragraph{DrongoNet-Micro} ($n_\mathrm{mels}{=}16$, 919 parameters) yields $\tau{=}0.35$, with per-seed recall 97.88\% / 98.32\% / 98.24\% / 98.44\% / 98.64\% (seeds 42/100/786/7/1234; mean 98.30\%) at 83--86\% precision.

\paragraph{DrongoNet-Edge} ($n_\mathrm{mels}{=}80$, ${\approx}25$\,K parameters) yields $\tau{=}0.425$, a single threshold across all five seeds, with per-seed recall 99.24\% / 99.12\% / 98.84\% / 98.80\% / 99.04\% (mean 99.01\%) at 97.29--98.80\% precision. A single $\tau$ across seeds simplifies deployment---practitioners can calibrate once rather than per checkpoint.

Both operating points are summarised in Table~\ref{tab:threshold_comparison}, with the full recall--precision sweep in Figure~\ref{fig:threshold_analysis}A and the per-threshold breakdown in Table~\ref{tab:perf_table}.

\begin{table*}[t]
	\centering
	\caption{Performance at operating decision thresholds (mean\,$\pm$\,std across five seeds)}
	\label{tab:threshold_comparison}
	\begin{tabular}{@{}llccccc@{}}
		\toprule
		Variant & $\tau_{\text{op}}$ & Recall & Precision & Specificity & F$_1$ & F$_2$ \\
		\midrule
		DrongoNet-Micro & \textbf{0.35} & 0.983$\pm$0.003 & 0.847$\pm$0.012 & 0.823$\pm$0.016 & 0.910$\pm$0.007 & 0.953$\pm$0.003 \\
		DrongoNet-Edge  & 0.425 & 0.990$\pm$0.002 & 0.984$\pm$0.006 & 0.984$\pm$0.006 & 0.987$\pm$0.002 & 0.989$\pm$0.001 \\
		\bottomrule
	\end{tabular}
	\smallskip\\
	{\footnotesize Values are the five-seed mean\,$\pm$\,std at each operating threshold, selected by the mean-recall rule of Section~\ref{sec:math}. Micro mean recall $0.983$ (min-seed 97.88\%, max 98.64\%); Edge mean recall $0.990$ (min-seed 98.80\%, max 99.24\%). Both clear their design floors (${\geq}98\%$ / ${\geq}99\%$) on the mean.}
\end{table*}

The sensitivity--specificity trade-off (Figure~\ref{fig:threshold_analysis}C) shows Micro at $\tau{=}0.35$ passing ${\approx}18\%$ of non-bird clips downstream. This is acceptable for a gatekeeper, where missed vocalisations are the primary failure mode, whereas Edge achieves near-symmetric recall and specificity (${\approx}98$--$99\%$ each).


\begin{figure}[!t]
	\centering
	\includegraphics[width=\textwidth]{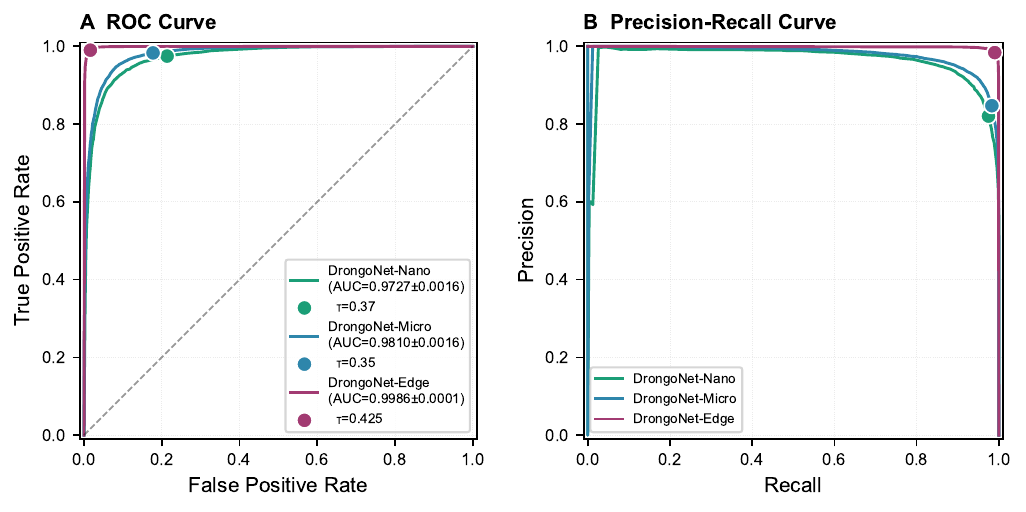}
	\caption{%
		\textbf{Discrimination performance of DrongoNet-Micro and DrongoNet-Edge.}
		Mean across five seeds; shaded bands $\pm$1\,std. Markers indicate operating
		points ($\tau{=}0.35$ for Micro; $\tau{=}0.425$ for Edge, uniform across
		all five seeds).
		\textbf{(A)}~ROC curves; dashed diagonal is chance. AUC annotated per variant.
		\textbf{(B)}~Precision--Recall curves with iso-F$_1$ contours.
	}
	\label{fig:roc_pr}
\end{figure}

\subsection{Discrimination Ability}

Figure~\ref{fig:roc_pr} shows ROC and Precision--Recall curves aggregated over five seeds. Seed-to-seed variation, visible in the shaded bands, is negligible for Edge and small for Micro---training is stable across initialisations.

\textbf{Edge} reaches AUC\,$=$\,0.9991\,$\pm$\,0.0002, near the ceiling on SEABAD, with the operating point sitting in the high-TPR, low-FPR corner of the ROC. \textbf{Micro} reaches AUC\,$=$\,0.9810\,$\pm$\,0.0016. The meaningful reference is not the DrongoNet-Edge ceiling but the retrained TinyChirp baseline ($0.9815\,\pm\,0.0008$, 25{,}558 params, $n_\mathrm{mels}{=}80$), since the original TinyChirp weights transfer only at chance. Against that baseline, Micro reaches near-identical discrimination (within 0.1\,pp) at $28\times$ fewer parameters, so efficiency rather than raw AUC is the decisive advantage, while Edge improves by 1.8\,pp at essentially the same parameter budget.

We tested both gaps for statistical significance (Welch's $t$-test and Mann--Whitney $U$ across the five/three per-seed AUCs, unpaired since the two architectures are trained independently): the Micro--TinyChirp difference is \emph{not} significant ($t{=}{-}0.57$, $p{=}0.59$; $U$-test $p{=}0.79$), confirming the ``near-identical'' claim rather than a favourable rounding; the Edge--TinyChirp gain \emph{is} significant ($t{=}39.5$, $p{<}0.001$; $U$-test $p{=}0.036$). Given $n{\leq}5$ seeds per arm these tests have limited power and are reported as a directional check, not a substitute for the effect sizes above.

The PR curves in Figure~\ref{fig:roc_pr}B make the gatekeeper trade-off explicit. Edge sustains ${\geq}98\%$ precision at 99.0\% recall, suitable for direct deployment. Micro operates at 84.7\% precision (roughly one false positive per five true positives)---acceptable for an MCU gatekeeper where a downstream classifier filters residual false alarms but not for stand-alone reporting.


\begin{figure}[htbp]
	\centering
	\includegraphics[width=\textwidth]{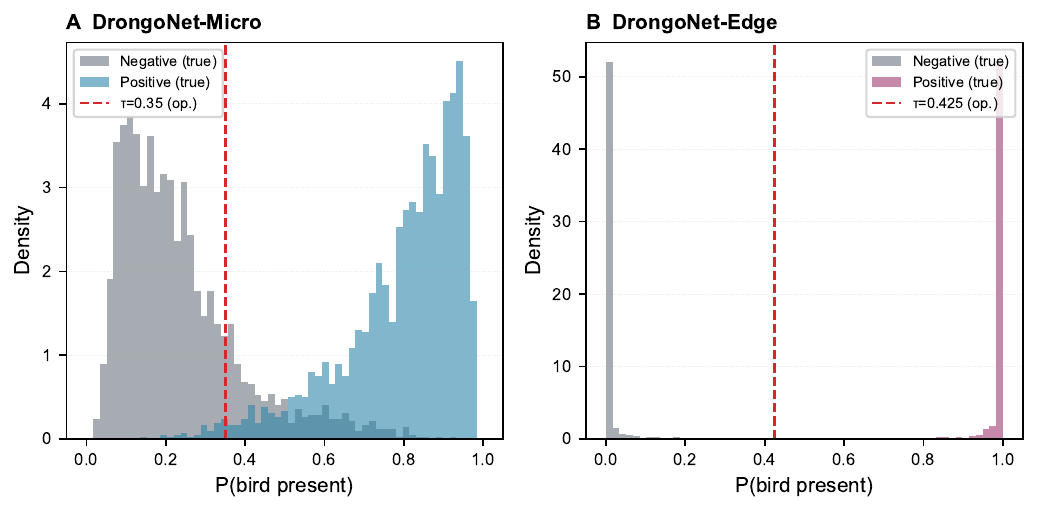}
	\caption{%
		\textbf{Predicted probability distributions for DrongoNet-Micro and DrongoNet-Edge.}
		\textbf{(A)} Normalised histograms by true class (positive: bird vocalisation;
		negative: non-bird). Dashed vertical lines mark the operating threshold
		($\tau{=}0.35$ for Micro; $\tau{=}0.425$ for Edge).
		\textbf{(B)} Cumulative distribution functions; thresholds fall at the
		${\geq}98\%$ positive-class quantile for both variants.
	}
	\label{fig:prob_dist}
\end{figure}

\subsection{Probability Distribution Analysis}

Figure~\ref{fig:prob_dist} shows predicted probability distributions stratified by true class. Both variants exhibit bimodal positive--negative separation, but with markedly different geometry. Focal-loss-trained Micro produces a compressed positive-class mode concentrated at lower probabilities, consistent with the calibrated $\tau{=}0.35$ operating point, while Edge produces a sharper, better-separated distribution that supports its $\tau{=}0.425$ operating point.

The downstream consequence is visible in the confusion-matrix counts at the operating threshold. Micro yields approximately 2{,}458 TP, 42 FN, 444 FP, and 2{,}056 TN per 5{,}000 test clips (means over seeds), an 18\% false-alarm rate that is acceptable for a gatekeeper stage where a downstream classifier filters the residual false positives. Edge yields ${\approx}2{,}475$ TP, 25 FN, 41 FP, and 2{,}459 TN: roughly $10\times$ fewer false positives than Micro, which makes it suitable for direct deployment scenarios with human review of flagged segments.


\subsection{Comprehensive Performance Summary}

Table~\ref{tab:deployment_comparison} consolidates the preceding subsections into a single side-by-side view: all three DrongoNet variants (Nano, Micro, Edge) against the retrained TinyChirp baseline, with AUC, recall at the operating threshold, model size, parameter count, and inference latency reported on a common basis. The recall row uses each variant's operating threshold from Section~\ref{sec:threshold}: Micro $\tau{=}0.35$ and Edge $\tau{=}0.425$ meet their $\geq$98\%/$\geq$99\% mean-recall floors, while Nano, which has no formal floor, uses $\tau{=}0.37$, tuned so every seed clears 97\% (mean 97.5\%, min-seed 97.0\%).

Two latency rows appear because they measure different things and must not be compared across rows. The \emph{host-CPU} row is a like-for-like x86-64 TFLite INT8 inference benchmark (1000-call mean, mel preprocessing excluded). The \emph{on-device} row places all four models on the same deployment target---the AudioMoth-class Cortex-M4F @ 48\,MHz---so its columns are directly comparable.

The TinyChirp baseline is Huang's directly measured 2386\,ms ($=$ 1980\,ms mel $+$ 406\,ms inference) on the NRF52840 Cortex-M4 \citep{huang2024tinychirp}, an M4-class core comparable to the AudioMoth M4F. The Nano and Micro figures (${\approx}370$\,ms and ${\approx}480$\,ms) are \emph{projected} for the AudioMoth M4F as sparse-filterbank mel (${\approx}162$\,ms at $n_\mathrm{fft}{=}512$ for Nano, ${\approx}273$\,ms at $n_\mathrm{fft}{=}1024$ for Micro; \S\ref{sec:deployment}) plus ${\approx}210$\,ms inference (the Portenta H7's 21\,ms scaled by the $10\times$ clock ratio, a lower bound since the M4F has narrower SIMD/MAC than the M7, and identical for Nano and Micro whose Portenta inference is the same 21\,ms).

Edge is marked N/A because it cannot run on AudioMoth at all---its 290\,kB tensor arena needs external SDRAM---and belongs to the Linux-SBC tier; its measured Portenta H7 Cortex-M7 pipeline is ${\approx}435$\,ms (Table~\ref{tab:inference_latency}). On this shared basis the ordering is unambiguous: the DrongoNet MCU variants are roughly $5$--$6\times$ faster than the TinyChirp baseline, and all deployable variants stay well inside the 3\,s analysis window.

\begin{table*}[htbp]
	\centering
	\caption{Performance comparison of model variants (TFLite INT8-quantised). AUC: mean\,$\pm$\,std across five seeds (42, 100, 786, 7, 1234) for DrongoNet, three (42, 100, 786) for the baseline. The AudioMoth-M4F latency row and its projection basis are described in the text; Edge is N/A there (needs SDRAM). Operating thresholds and recall floors are set in Section~\ref{sec:threshold}.}
	\label{tab:deployment_comparison}
	\smallskip
	\begin{tabular}{@{}lcccc@{}}
		\toprule
		\textbf{Metric} & \textbf{Baseline} & \textbf{DrongoNet-Nano} & \textbf{DrongoNet-Micro} & \textbf{DrongoNet-Edge} \\
		\midrule
		\textbf{AUC} & 0.9815 $\pm$ 0.0008 & $0.9727 \pm 0.0016$ & $0.9810 \pm 0.0016$ & $0.9991 \pm 0.0002$ \\
		\textbf{Recall} (op.\ threshold) & --- & $97.5\%$ @ $\tau{=}0.37$ & $98.3\%$ @ $\tau{=}0.35$ & $99.0\%$ @ $\tau{=}0.425$ \\
		\textbf{Model Size (INT8)} & 28.61\,kB & \textbf{5.09\,kB} & \textbf{6.26\,kB} & 33.06\,kB \\
		\textbf{Params} & 25,558 & \textbf{763} & \textbf{919} & \textbf{25,890} \\
		\textbf{Latency}, host CPU (infer.) & 0.34\,ms & 0.09\,ms & 0.10\,ms & 1.13\,ms \\
		\textbf{Latency}, AudioMoth M4F (mel+infer) & 2386\,ms & ${\approx}370$\,ms & ${\approx}480$\,ms & N/A \\
		\textbf{$n_\mathrm{mels}$} & 80 & 16 & 16 & 80 \\
		\textbf{$n_\mathrm{fft}$} & 1024 & 512 & 1024 & 1024 \\
		\textbf{Target Platform} & ARM Cortex-M4 & ARM Cortex-M4 & ARM Cortex-M4 & Linux SBC \\
		\bottomrule
	\end{tabular}
\end{table*}

\section{Embedded Deployment Evaluation}
\label{sec:deployment}

The deployment question for a TinyML bird detector is not ``does the model run?'' but ``does it leave the sensor in the field longer than the trigger it replaces?'' \citet{benhammadi2026battery} make this concrete for the Cortex-M4F AudioMoth: against a 210-hour indiscriminate-recording baseline, the AudioMoth-native Goertzel trigger yields no net gain (it saves SD writes but its 71\% precision floods the card with false positives), a mel-based DNN trigger gains only 6 hours (216 h), while their lightest detector (a Gabor filterbank operating on raw waveform) gains 31 hours (241 h)~\citep{benhammadi2026battery}.

The lesson for any mel-domain detector targeting AudioMoth is therefore twofold: (i) the mel front-end must cost no more than the convolutions it feeds, and (ii) per-segment precision must be high enough that fewer writes actually translate into longer battery life. This section establishes that DrongoNet-Micro meets both bars, using on-device measurements rather than host-CPU extrapolations.

\subsection{Mel Preprocessing on ARM Cortex-M}

On AudioMoth-class hardware the trained model is not the bottleneck---mel preprocessing is. A naive dense filterbank would consume tens of millions of cycles per frame and dominate the energy budget; the only way to fit a mel-input architecture inside the AudioMoth thermal and current envelope (32\,kB SRAM, ${<}10$\,mA average~\citep{benhammadi2026battery}) is to push every step of preprocessing through CMSIS-DSP~\citep{lai2018cmsisnn}. Table~\ref{tab:melcalc} reports per-frame cycle counts for an 80-mel, 184-frame spectrogram so configured.

\begin{table}
	\centering
	\caption{Mel spectrogram computation estimates on AudioMoth (Silicon Labs EFM32 Wonder Gecko, Cortex-M4F @ 48\,MHz) using CMSIS-DSP kernels~\citep{lai2018cmsisnn}. Energy assumes the 3.3\,V supply and 9.20\,mA always-on (audio acquisition) current measured by \citet{benhammadi2026battery}.}
	\label{tab:melcalc}
	\smallskip
	\begin{tabular}{@{}lrr@{}}
		\toprule
		\textbf{Step} & \textbf{$n_\mathrm{fft}{=}512$} & \textbf{$n_\mathrm{fft}{=}1024$} \\
		\midrule
		Windowing & 1,536 & 3,072 \\
		Real FFT & 30,457 & 55,538 \\
		Power spectrum & 771 & 1,539 \\
		Mel filterbank (sparse) & $\approx$1,500 & $\approx$3,200 \\
		Logarithm (80 bins) & 8,000 & 8,000 \\
		\midrule
		Total per frame & $\approx$42,264 & $\approx$71,349 \\
		Total (184 frames) & $\approx$7.78\,M & $\approx$13.13\,M \\
		\midrule
		\textbf{Overall latency} & \textbf{$\approx$162\,ms} & \textbf{$\approx$273\,ms} \\
		\textbf{Energy consumed} & \textbf{$\approx$4.9\,mJ} & \textbf{$\approx$8.3\,mJ} \\
		\bottomrule
	\end{tabular}
\end{table}

The single optimisation that keeps mel preprocessing tractable on M4F is the sparse filterbank: processing only the non-zero triangular weights cuts cycles by over $95\%$, holding total preprocessing for a 3-second clip to ${\approx}162$\,ms ($n_\mathrm{fft}{=}512$, Nano) or ${\approx}273$\,ms ($n_\mathrm{fft}{=}1024$, Micro/Edge). The per-window energy cost (${\approx}5$--$8$\,mJ) remains a small fraction of the ${\approx}90$\,mJ the always-on front-end already spends acquiring each 3-second window, and is dwarfed by the SD write it gates. In \citet{benhammadi2026battery}'s on-AudioMoth measurements, the Goertzel trigger averages 12.39\,mA, while their Mel-DNN trigger averages 12.04\,mA at recall $\sim$0.9; mel preprocessing is therefore not the line item that decides battery life.

What decides battery life is \emph{precision}: every false positive Goertzel raises forces a 0.8\,s SD write that costs an order of magnitude more energy than the trigger itself, which is why Goertzel's 0.71 precision~\citep{benhammadi2026battery} erases its compute savings. The energy case for a mel-domain trigger on AudioMoth therefore rests on combining (i) a sparse-filterbank front-end that keeps preprocessing within a few mA of the always-on baseline, and (ii) a detection head whose precision is high enough to leave fewer writes than Goertzel. DrongoNet-Micro is engineered to do both: at $\tau{=}0.35$ it achieves $98.3\%$ recall with FPR $=0.18$ (Section~\ref{sec:threshold}), translating directly into the storage-day projections discussed in Section~\ref{sec:discussion}.

\subsection{Cross-Platform Inference Latency and Power}

To establish that the model itself does not blow the energy budget, we report on-device measurements from a Portenta H7 (STM32H747XI, Cortex-M7 @ 480\,MHz) running TFLite for Microcontrollers under bare-metal Mbed OS, instrumented with an INA219 power monitor and CMSIS-DSP mel preprocessing identical to the AudioMoth path. The Portenta is not an AudioMoth, but it is the closest microcontroller-class platform on which all three DrongoNet variants (including the SDRAM-resident Edge) will run; the M4F AudioMoth is roughly $10\times$ slower per cycle but draws roughly $10\times$ less current, so per-window energy scales similarly. A host-CPU TFLite INT8 row is included as a relative-throughput reference, not as a deployment claim (Table~\ref{tab:inference_latency}).

\begin{table}[ht]
	\centering
	\caption{DrongoNet on-device latency, memory, and power, measured on a Portenta H7 (STM32H747XI Cortex-M7 @ 480\,MHz, TFLite for Microcontrollers, bare-metal Mbed OS, INA219 at 5.0\,V). Host-CPU rows are an x86-64 TFLite INT8 throughput reference (1000-call mean) and are \emph{not} on-MCU measurements.}
	\label{tab:inference_latency}
	\small
	\begin{tabular}{@{}lrrr@{}}
		\toprule
		\textbf{Metric} & \textbf{Nano} & \textbf{Micro} & \textbf{Edge} \\
		\midrule
		Mel preprocessing (Portenta H7)  & 32\,ms     & 32\,ms     & 36--37\,ms     \\
		Inference (Portenta H7)          & 21\,ms     & 21\,ms     & 398\,ms        \\
		Per-window pipeline (mel+infer)  & 53\,ms     & 53\,ms     & $\approx$435\,ms \\
		Compute duty cycle (3\,s window) & 1.8\%      & 1.8\%      & 14.5\%         \\
		\midrule
		TFLM tensor arena (used)         & 23.2\,kB   & 23.0\,kB   & 290\,kB (SDRAM) \\
		Min.\ viable arena ($\times 1.12$) & 25.9\,kB & 25.8\,kB   & 324.8\,kB       \\
		Model flash footprint            & 5.09\,kB   & 6.26\,kB   & 33.06\,kB       \\
		\midrule
		Steady-state power (active)      & 78--144\,mW & 184--212\,mW & 440--466\,mW \\
		Inference (host CPU, x86-64)     & 0.09\,ms   & 0.10\,ms   & 1.13\,ms        \\
		\bottomrule
	\end{tabular}
	\smallskip\\
	{\footnotesize Portenta H7 numbers come from a memory-profiling firmware build with INA219 sampling per pipeline stage at 5.0\,V. Edge's 290\,kB arena lives in external SDRAM, which is what drives its elevated steady-state current (90--96\,mA). Micro/Nano arenas live in on-chip RAM and fit comfortably within the 32\,kB SRAM ceiling that \citet{benhammadi2026battery} identify for AudioMoth-class deployment.}
\end{table}

\textbf{Three observations emerge from the on-device evaluation.}

\emph{First, DrongoNet-Micro is duty-cycle dominated rather than compute dominated.} On the Portenta H7, mel-spectrogram extraction and inference for a single 3-second window require $32 + 21 = 53$\,ms, corresponding to a compute duty cycle of $53/3000 = 1.8\%$. More than 98\% of wall-clock time is therefore spent sampling audio or waiting for the next analysis window. Under these conditions, overall system energy consumption is influenced more by sampling and idle behaviour than by neural-network execution.

\emph{Second, the memory footprint is compatible with AudioMoth-class hardware.} The Portenta H7 firmware reports a TFLite Micro tensor arena of 23.02\,kB for DrongoNet-Micro, including a 2.94\,kB input tensor and a 2-byte output tensor for the two-class softmax. The recommended minimum arena size is 25.8\,kB, which remains within the 32\,kB SRAM budget reported for AudioMoth deployments by \citet{benhammadi2026battery}. DrongoNet-Edge requires a 290\,kB tensor arena and therefore relies on external SDRAM, making it more appropriate for SBC-class platforms such as Raspberry Pi systems or Portenta H7 configurations equipped with external memory.

\emph{Third, the measured power consumption is consistent with AudioMoth-class deployment requirements.} On the Portenta H7 operating at 5.0\,V, DrongoNet-Micro draws 36--42\,mA during the active processing interval, corresponding to 184--212\,mW. Because computation occupies only 1.8\% of each analysis cycle, the average current is substantially lower when idle periods are included. When scaled against the lower-power Cortex-M4F platform used by AudioMoth, these measurements are consistent with the sub-10\,mA average-current target identified by \citet{benhammadi2026battery} for long-term TinyML deployments. Their AudioMoth Mel detector reports an average current of 12.04\,mA at approximately 0.90 recall, whereas DrongoNet-Micro achieves approximately 0.983 recall (Section~\ref{sec:threshold}) while remaining within the same deployment class.

Taken together, the results indicate distinct deployment roles for the three model variants. DrongoNet-Nano and DrongoNet-Micro are intended for AudioMoth-class trigger applications, where SRAM, flash capacity, and average current consumption are the primary constraints. Nano targets the smallest memory budgets, while Micro provides the best overall trade-off between footprint and detection performance. DrongoNet-Edge serves as a higher-capacity reference design for platforms with external memory, illustrating the accuracy attainable when memory constraints are relaxed.

\section{Discussion}
\label{sec:discussion}

\subsection{Model Performance and Generalisation}

A natural concern is whether DrongoNet's design overfits the SEABAD benchmark. Three transfer experiments argue otherwise, each probing a different axis of generalisation and each carrying a direct deployment consequence.

\paragraph{Cross-region, cross-species retraining.} When retrained from scratch on the TinyChirp Corn Bunting corpus~\citep{huang2024tinychirp}, a different region, species, and ecological niche, DrongoNet-Micro reaches $0.9757 \pm 0.0085$ AUC at $28\times$ fewer parameters than the purpose-built TinyChirp CNN-Mel, Edge reaches $0.9997 \pm 0.0004$, and Nano reaches $0.9664 \pm 0.0084$. The architecture exhibits stable per-dataset AUC across species, meaning the same network topology can be retrained on regional data without architectural modification and achieves competitive results. Field operators wishing to deploy in a new region must retrain on representative acoustic data from that region and then optimise the decision threshold.

\paragraph{Against a general-purpose foundation model.} BirdNET GLOBAL 6K (51.7\,MB~\citep{kahl2021birdnet}) zero-shot on SEABAD attains only $0.940$ AUC and ${\leq}82\%$ recall---below DrongoNet-Micro at ${\sim}8300\times$ smaller footprint. The binary gatekeeper task does not benefit from BirdNET's species-level richness; instead, a small task-specific detector trained on the right corpus dominates a large general-purpose one. This is the case for maintaining the gatekeeper-then-classifier separation rather than collapsing both into a single large model.

\paragraph{In-domain efficiency on a temperate benchmark (DCASE-2018).}
A third test retrains DrongoNet from scratch on the DCASE-2018 Bird Audio Detection development set (freefield1010\,+\,Warblr, 10-second clips), applying the 3-second model via six overlapping windows with max aggregation, and compares against bulbul~\citep{grill2017two}. In-domain (stratified split, seeds 123/456/789; Table~\ref{tab:dcase_bulbul}), Edge reaches $0.938 \pm 0.010$ AUC---within 0.4\,pp of bulbul's in-protocol $0.942$ at $\mathbf{14.4{\times}}$ \textbf{fewer parameters} and with no data augmentation. Micro ($0.847$) and Nano ($0.821$) trail Edge by ${\sim}9$--$12$\,pp at $406$--$489\times$ fewer parameters than bulbul, locating both on a clean accuracy--size frontier.

The architecture retrains successfully on a different clip length and temperate soundscape without modification to the network or training recipe, demonstrating architectural robustness across acoustic domains when given representative in-domain data. Among the three transfer experiments, the DCASE result is the most directly comparable to published prior work: a temperate benchmark with a citable peer baseline where parameter efficiency is the decisive advantage.

\paragraph{Limits of zero-shot cross-corpus transfer.}
The right-hand column of Table~\ref{tab:dcase_bulbul} reports the same DCASE-trained models evaluated zero-shot on BirdVox-DCASE-20k, without domain adaptation. Edge holds at $0.670$; Micro ($0.519$) and Nano ($0.468$) collapse toward chance. Bulbul, measured in the identical pipeline, falls from $0.942$ in-domain to $0.732/0.749$ (without/with augmentation) cross-corpus---a ${\sim}19$--$21$\,pp drop on the same corpus shift that costs DrongoNet 27--35\,pp. Augmentation helps only at capacity (bulbul $+1.7$\,pp, Nano $+7.9$\,pp) and regresses Micro and Edge, confirming that at MCU scale there is no spare capacity to absorb the acoustic-environment shift between two temperate corpora.

The size of the gap is real and is the cost of working at 763--25{,}890 parameters. DrongoNet does not transfer zero-shot across acoustic domains and is not intended to: field deployment in a new region (Amazonian, African, or Oceanian rainforests; non-Southeast-Asian temperate corpora) requires retraining on regional data of the kind we collected for SEABAD, after which the in-domain efficiency results above predict the achievable AUC budget. Section~\ref{sec:limitations} folds this into the deployment scope.

\begin{table}[ht]
	\centering
	\caption{DrongoNet on DCASE-2018 BAD versus bulbul~\citep{grill2017two}, all measured in one pipeline (protocol in text). \emph{In-domain}: freefield1010\,+\,Warblr; \emph{cross-corpus}: tested zero-shot on BirdVox-DCASE-20k (5 seeds). bulbul values are our in-protocol re-measurements, not its paper-reported 0.96\,/\,0.887.}
	\label{tab:dcase_bulbul}
	\small
	\begin{tabular}{lrcc}
		\toprule
		\textbf{System} & \textbf{Params} & \textbf{In-domain AUC} & \textbf{Cross-corpus AUC} \\
		\midrule
		bulbul~\citep{grill2017two}  & 373,169 & $0.942\,\pm\,0.005$       & 0.732 / 0.749$^\ast$ \\
		DrongoNet-Edge (80 mel)      & 25,890  & $0.938\,\pm\,0.010$       & 0.670 \\
		DrongoNet-Micro (16 mel)     & 919     & $0.847\,\pm\,0.011$       & 0.519 \\
		DrongoNet-Nano (16 mel)      & 763     & $0.821\,\pm\,0.002$       & 0.468 \\
		\bottomrule
	\end{tabular}
	\\[2pt]
	{\footnotesize $^\ast$bulbul cross-corpus AUC without\,/\,with data augmentation, measured in our pipeline (mean over 5 seeds); un-augmented DrongoNet cross-corpus values shown for the other rows.}
\end{table}

\subsection{Implications for Autonomous Acoustic Monitoring}
\label{sec:implications}

The deployment question raised in Section~\ref{sec:deployment} can now be closed with numbers. DrongoNet-Micro is engineered as a drop-in replacement for the AudioMoth Goertzel frequency-energy trigger~\citep{benhammadi2026battery}, and its on-card storage and battery-life implications follow directly from its measured recall--FPR profile combined with Benhammadi's renewal-reward energy model.

\paragraph{Storage on a 32\,GB SD card.}
At tropical bird-segment prevalence $\alpha=0.10$, Micro at $\tau{=}0.35$ (recall ${\approx}98\%$, FPR\,$=$\,0.18) records 25.8\% of segments versus Goertzel's 41.7\%~\citep{benhammadi2026battery}, extending a 32\,GB card from ${\approx}28$ to ${\approx}45$ days of continuous monitoring (Table~\ref{tab:storage}). At the higher-precision $\tau{=}0.50$ (recall ${\approx}94\%$), the same card lasts ${\approx}73$ days. These projections are conservative for tropical soundscapes, where Goertzel's false-positive rate is expected to exceed its temperate-survey baseline~\citep{benhammadi2026battery}---so the real-world storage gain over Goertzel should be larger than the table suggests.

\begin{table}[ht]
	\centering
	\caption{Storage on a 32\,GB SD card at tropical prevalence $\alpha=0.10$ (16\,kHz mono, 3-second clips, ${\approx}115$\,MB/h continuous). Hit rate is the fraction of windows written, $p=\alpha\cdot\text{Recall}+(1{-}\alpha)\cdot\text{FPR}$; days $=$ Goertzel's ${\approx}28$-day baseline scaled by the inverse hit-rate ratio. Goertzel and DrongoNet-Micro operate at matched high recall; the storage difference is driven by precision. Reference row from \citet{benhammadi2026battery}.}
	\label{tab:storage}
	\small
	\begin{tabular}{@{}lcccc@{}}
		\toprule
		\textbf{Detector} & \textbf{Recall} & \textbf{Precision} & \textbf{Hit rate} & \textbf{32\,GB lasts} \\
		\midrule
		Goertzel (AudioMoth)              & 0.90  & 0.71  & 41.7\% & ${\approx}28$\,d \\
		\midrule
		DrongoNet-Micro ($\tau{=}0.35$)   & 0.983 & 0.847 & 25.8\% & ${\approx}45$\,d \\
		DrongoNet-Micro ($\tau{=}0.50$)   & 0.938 & 0.927 & 16.0\% & ${\approx}73$\,d \\
		\bottomrule
	\end{tabular}
\end{table}

\paragraph{Battery lifetime via the Benhammadi model.}
We project DrongoNet-Micro's AudioMoth battery lifetime by applying the expected-current model of \citet{benhammadi2026battery} (their Equation A.14):

\begin{equation}
\label{eq:iavg}
I_\textrm{avg} = \frac{I_\textrm{always\text{-}on} \cdot T_\textrm{always\text{-}on} + p \cdot I_\textrm{on\text{-}detect} \cdot T_\textrm{on\text{-}detect}}{T_\textrm{always\text{-}on} + p \cdot T_\textrm{on\text{-}detect}},
\end{equation}

\noindent with $p = \alpha \cdot \text{Recall} + (1{-}\alpha) \cdot \text{FPR}$ as the per-window write probability. Because DrongoNet-Micro shares its CMSIS-DSP mel front-end with Benhammadi's Mel detector, we adopt their measured Mel-detector coefficients ($I_{always\text{-}on}=11.45$\,mA, $T_{always\text{-}on}=1500$\,ms, $I_{on\text{-}detect}=18.20$\,mA, $T_{on\text{-}detect}=1650$\,ms, $C=2600$\,mAh) and substitute the DrongoNet-Micro hit rate at each operating point (Table~\ref{tab:battery_projection}). Reference values for indiscriminate recording (210\,h), Goertzel (210\,h, no gain due to its 0.71 precision), Mel-DNN (216\,h at $p=0.09$), and Gabor (241\,h, the +31\,h headline) are reproduced from \citet{benhammadi2026battery} Table~4.

\begin{table}
	\centering
	\caption{Projected on-AudioMoth battery lifetime for DrongoNet-Micro at three field prevalences, using Equation~\ref{eq:iavg} with \citet{benhammadi2026battery}'s Mel-detector electrical coefficients (mel preprocessing shared). Reference rows reproduce \citet{benhammadi2026battery} Table~4 directly. $p = \alpha \cdot \text{Recall} + (1{-}\alpha) \cdot \text{FPR}$ is the per-window write probability; $L(C)$ is battery lifetime at $C{=}2600$\,mAh, $5$\,V.}
	\label{tab:battery_projection}
	\small
	\begin{tabular}{@{}llccrrr@{}}
		\toprule
		\textbf{Detector} & \textbf{Setting} & \textbf{Rec.} & \textbf{Prec.} & \textbf{$p$} & \textbf{$I_{avg}$ (mA)} & \textbf{$L(C)$ (h)} \\
		\midrule
		Indiscriminate (reference)        & ---        & ---  & ---  & 1.00 & 12.39$^\ast$ & 210$^\ast$ \\
		Goertzel (reference)              & temperate  & 0.90 & 0.71 & 1.00 & 12.39$^\ast$ & 210$^\ast$ \\
		Mel-DNN (reference)               & temperate  & 0.90 & 0.96 & 0.09 & 12.04$^\ast$ & 216$^\ast$ \\
		Gabor (reference)                 & temperate  & 0.90 & 0.93 & 0.14 & 10.97$^\ast$ & 241$^\ast$ \\
		\midrule
		DrongoNet-Micro                   & $\alpha=0.10$, $\tau=0.35$ & 0.98 & 0.85 & 0.258 & ${\approx}12.9$ & ${\approx}201$ \\
		DrongoNet-Micro                   & $\alpha=0.10$, $\tau=0.50$ & 0.94 & 0.93 & 0.160 & ${\approx}12.5$ & ${\approx}209$ \\
		DrongoNet-Micro                   & $\alpha=0.05$, $\tau=0.35$ & 0.98 & 0.85 & 0.218 & ${\approx}12.8$ & ${\approx}204$ \\
		DrongoNet-Micro                   & $\alpha=0.05$, $\tau=0.50$ & 0.94 & 0.93 & 0.117 & ${\approx}12.2$ & ${\approx}213$ \\
		DrongoNet-Micro                   & $\alpha=0.01$, $\tau=0.50$ & 0.94 & 0.93 & 0.083 & ${\approx}12.0$ & ${\approx}216$ \\
		\bottomrule
	\end{tabular}
	\\[2pt]
	{\footnotesize $^\ast$Directly measured on AudioMoth by \citet{benhammadi2026battery} (their Table~4); reference-detector recall/precision are their reported values at the recall-0.9 operating point (Gabor 0.93, Mel-DNN 0.96, Goertzel 0.71). DrongoNet-Micro rows are model projections using their Mel-detector electrical coefficients with DrongoNet-Micro's measured recall and precision; an on-AudioMoth firmware port and INA219 validation are stated as future work. At matched recall the reference triggers cluster near 0.9, but only Goertzel's precision (0.71) is low enough to erase its write savings---the reason it yields no lifetime gain over indiscriminate recording.}
\end{table}

Two conclusions follow. First, at the recall-prioritised operating point ($\tau{=}0.35$, recall $98.3\%$), Micro at $\alpha=0.10$ projects to ${\approx}201$\,h---below the 210\,h always-on baseline. This is the honest cost of pursuing very high recall in a moderately bird-active soundscape: the model triggers a write on roughly one-quarter of all 3-second windows, and the per-window SD-write energy then dominates. At this recall-first operating point ($\tau{=}0.35$), the primary deployment benefit is storage extension (45 vs.\ 28 days, ${\approx}61\%$ gain), while battery life (201\,h) remains below the 210\,h always-on baseline. This recall-first operating mode suits operators prioritising detection sensitivity over battery longevity, yielding an ${\approx}8.3$\,pp recall lift over Benhammadi's Mel-DNN baseline.

Second, at the higher-precision $\tau{=}0.50$ operating point and at lower prevalence, which characterises most tropical and rainforest surveys with a single target species or species group, DrongoNet-Micro projects to ${\approx}213$\,h at $\alpha=0.05$ and ${\approx}216$\,h at $\alpha=0.01$, equalling \citet{benhammadi2026battery}'s on-AudioMoth Mel detector (216\,h). It does so while operating at $\geq{}93\%$ recall, above their Mel detector's pre-set 90\%, and with substantially higher precision than Goertzel. The architectural takeaway is that DrongoNet-Micro converts most of Benhammadi's measured Mel-trigger battery envelope into an additional 3--8\,pp of recall, rather than into more hours; field operators can choose between recall-first and lifetime-first by toggling a single threshold, no retraining required.

The Gabor detector's $+31$\,h headline (241\,h vs 210\,h) remains the on-AudioMoth lifetime record~\citep{benhammadi2026battery}; the present results do not contest it. They establish a different point: a CNN-based trigger can be made small enough, sparse-mel enough, and precise enough to deliver \emph{both} an improved storage envelope \emph{and} battery life within ${\sim}10$\,h of Gabor's, while delivering substantially higher recall in the prevalence regime where most tropical PAM surveys actually operate.

\subsection{Deployment Guide: Choosing the Right Variant and Threshold}

The variant and threshold are selected jointly from two field-side inputs: expected bird-segment prevalence $\alpha$ and the operator's tolerance for downstream false positives. SEABAD is class-balanced for training, but tropical field deployments are sparse---prevalence is typically 1:20 to 1:100 \citep{lostanlen2018pervasive}. Propagating test-set TPR/FPR to those prevalences via Bayes' rule yields the field-precision matrix in Table~\ref{tab:prevalence_analysis}, which doubles as a deployment-selection guide.

\begin{table}[ht]
	\centering
	\caption{Predicted field precision (PPV, \%) at representative bird-segment prevalence, derived from test-set TPR/FPR via Bayes' rule. Micro $\tau{=}0.35$: TPR/FPR = 0.983/0.178; Micro $\tau{=}0.50$: 0.938/0.074; Edge ($\tau{=}0.425$, uniform): 0.990/0.016. A clip-level bootstrap (B=5000) confirms projections to within ${\pm}0.2$\,pp.}
	\label{tab:prevalence_analysis}
	\small
	\begin{tabular}{cccc}
		\toprule
		\textbf{Bird prevalence} & \textbf{Micro $\tau{=}0.35$} & \textbf{Micro $\tau{=}0.50$} & \textbf{Edge} \\
		\midrule
		1:1 (balanced test) & 85\% & 93\% & 98\% \\
		1:10                & 36\% & 56\% & 86\% \\
		1:20                & 22\% & 39\% & 75\% \\
		1:50                & 10\% & 20\% & 55\% \\
		1:100               &  5\% & 11\% & 38\% \\
		\bottomrule
	\end{tabular}
\end{table}

The matrix admits three operating regimes:

\emph{Moderate prevalence ($\alpha \geq 0.05$), MCU deployment.} DrongoNet-Micro at $\tau{=}0.35$ meets the design-floor recall (98.3\% mean; ${\geq}97.9\%$ per seed) at the cost of precision dropping to 22--36\% in the field at $\alpha \in \{1{:}10, 1{:}20\}$. This is the right setting when missed events are categorically more costly than false alarms---continuous community surveys, automated abundance estimation, alarm-call monitoring---and when a downstream classifier or human reviewer will handle the false positives. Per-seed recall is $97.88\%$, $98.32\%$, $98.24\%$, $98.44\%$, and $98.64\%$ (seeds 42/100/786/7/1234): the mean floor has measurable margin.

\emph{Lower prevalence or storage-constrained deployments, MCU.} DrongoNet-Micro at $\tau{=}0.50$ trades $\sim$4.5\,pp of recall for $\sim$2$\times$ better precision (39--56\% at $\alpha \in \{1{:}10, 1{:}20\}$). This is the recommended default when the SD card is the binding constraint or when battery lifetime parity with the Mel-DNN baseline matters more than catching every vocalisation (Table~\ref{tab:battery_projection}). Practitioners can raise $\tau$ further to $0.70$--$0.85$ for rare-event monitoring if they accept recall in the 97--98\% range (per Section~\ref{sec:deployment}).

\emph{Rare events ($\alpha \leq 0.02$), SBC deployment.} DrongoNet-Edge's order-of-magnitude lower FPR (0.016 versus Micro's 0.074 at $\tau{=}0.50$) preserves 38--55\% field precision even at $\alpha \in \{1{:}50, 1{:}100\}$---a regime where Micro precision collapses to single digits. Edge requires SDRAM-class memory (290\,kB arena) and cannot run on AudioMoth itself~\citep{benhammadi2026battery}; it is the appropriate choice for Raspberry-Pi-class triggers monitoring elusive single species. The penalty is power (440--466\,mW active versus Micro's 184--212\,mW) and a 15\% compute duty cycle in place of Micro's 1.8\%.

\subsection{Limitations and Future Directions}
\label{sec:limitations}

\paragraph{Zero-shot cross-corpus transfer is not claimed.}
As Table~\ref{tab:dcase_bulbul} shows, DrongoNet-Micro and -Nano drop to chance when evaluated zero-shot across acoustic domains (DCASE-2018 dev $\to$ BirdVox-DCASE-20k). At MCU-scale parameter budgets the network has no spare capacity to absorb environment shifts of this size, and the bulbul comparison shows that even a $14\times$ larger model loses 7--10\,pp under the same protocol. The deployment scope of this paper is therefore explicitly \emph{in-region}: DrongoNet should be retrained on representative regional data before fielding it in any soundscape materially different from SEABAD's Southeast Asian distribution. The in-domain DCASE result (\S\ref{sec:discussion}) characterises the AUC budget once that retraining is done.

\paragraph{Mel-domain augmentation does not substitute for capacity at MCU scale.}
A natural follow-up is whether the cross-corpus gap can be closed by stronger regularisation rather than by more parameters or target-domain data. We tested this with a controlled sweep of the full augmentation ladder---none, mixup~\citep{zhang2018mixup}, SpecAugment~\citep{park2019specaugment}, and the strongest stacked recipe (\emph{full}: SpecAugment\,$\to$\,frequency/time warp\,$\to$\,mixup)---applied to all three variants under one protocol: train on the DCASE-2018 development set (ff1010bird\,+\,warblrb10k), 50 epochs, six-window MAX aggregation, seeds 42/100/786/7/1234, evaluated zero-shot on BirdVox-DCASE-20k (Table~\ref{tab:aug_sweep}).

The operative variable is capacity, not augmentation. Un-augmented cross-corpus transfer rises monotonically with model size (Nano $<$ Micro $<$ Edge, Table~\ref{tab:aug_sweep}), with only Edge ($n_\mathrm{mels}{=}80$, $25{,}890$ params) approaching a usable ${\geq}0.70$ operating point while the two $16$-mel MCU variants sit near chance. Augmentation neither changes this ordering nor closes the gap: the strongest recipe helps only the smallest model (Nano $+7.9$\,pp) and \emph{regresses} both Micro ($-5.5$\,pp) and Edge ($-9.0$\,pp), eroding precisely the capacity headroom that makes un-augmented Edge the best transfer model---and for Edge every rung of the ladder falls below its no-augmentation baseline.

For Micro the same recipes also trade against fitting the training distribution (in-domain validation AUC $0.865\to0.803$), consistent with a 919-parameter model having no spare capacity to exploit.

This contrasts with bulbul's report that augmentation aids cross-corpus transfer at large scale ($+3$\,pp at $373$k parameters;~\citealp{grill2017two}), an order of magnitude above our tested range. At the sub-1\,kB budgets that define the MCU contribution, stacked mel-domain augmentation cannot substitute for capacity or target-domain data---so this strengthens rather than qualifies the in-region deployment scope: target-domain retraining is the appropriate remedy, with domain-adversarial training a complementary direction.

\begin{table}[ht]
	\centering
	\caption{Cross-corpus augmentation sweep: zero-shot AUC on BirdVox-DCASE-20k after training on DCASE-2018 dev (ff1010bird\,+\,warblrb10k), mean\,$\pm$\,std over seeds 42/100/786/7/1234 under a single protocol (10\,\% held-out validation, 50 epochs, six-window MAX aggregation). \emph{full} stacks SpecAugment, frequency/time warp, and mixup. The full ladder (none/mixup/specaug/full) is swept for all three variants. The final column reports the full\,$-$\,none difference in AUC percentage points (pp).}
	\label{tab:aug_sweep}
	\small
	\begin{tabular}{lrccccc}
		\toprule
		\textbf{Variant} & \textbf{Params} & \textbf{none} & \textbf{mixup} & \textbf{specaug} & \textbf{full} & \textbf{$\Delta$(full$-$none)} \\
		\midrule
		Nano  & 763      & $0.468\,{\pm}\,0.050$ & $0.475\,{\pm}\,0.018$ & $0.443\,{\pm}\,0.030$ & $0.547\,{\pm}\,0.055$ & $+7.9$\,pp \\
		Micro & 919      & $0.519\,{\pm}\,0.029$ & $0.514\,{\pm}\,0.030$ & $0.484\,{\pm}\,0.029$ & $0.465\,{\pm}\,0.031$ & $-5.5$\,pp \\
		Edge  & 25{,}890 & $0.670\,{\pm}\,0.030$ & $0.657\,{\pm}\,0.015$ & $0.646\,{\pm}\,0.021$ & $0.579\,{\pm}\,0.034$ & $-9.0$\,pp \\
		\bottomrule
	\end{tabular}
\end{table}

\paragraph{Knowledge distillation from a larger teacher does not improve Micro.}
Because the 6\,kB-scale architecture search is bounded by TFLite flatbuffer overhead rather than useful capacity, we tested whether the accuracy gap to DrongoNet-Edge could be closed on the training side instead---via knowledge distillation (KD), which changes only the training loss and adds zero inference cost. Following the WrenNet recipe \citep{ciapponi2025enabling}, we trained students with Micro's exact 919-parameter architecture against the trained Edge teacher (float AUC $0.9991$), both with and without SpecAugment \citep{park2019specaugment}, over seeds 42/100/786. Both variants \emph{underperformed} training Micro from scratch: KD reached $0.9756$ AUC and KD+SpecAugment $0.9719$, against the plain control's $0.9810$---a $0.5$--$1.4$\,pp deficit, with elevated INT8 degradation rather than the control's near-zero.

The likely cause is a teacher--student mismatch: the two see the same mel input at different resolutions ($n_\mathrm{mels}{=}80$ vs.\ $16$), so the teacher's soft targets encode frequency-scale detail the 16-mel student cannot represent, making them a poor training signal at this compression ratio. We therefore ship the from-scratch Micro and report this as a negative result to spare other MCU practitioners the same experiment.

\paragraph{Negative-corpus coverage and tropical confounders.}
Qualitative error analysis shows the residual errors are concentrated in characteristically tropical confounders: false negatives skew toward low-SNR and atypical vocalisations (alarm calls, subsongs), while false positives are dominated by insect stridulation, wind-induced vegetation rustle, and primate vocalisations---all underrepresented in the negative training corpus. SEABAD is also regionally scoped to Southeast Asia, so the cross-regional zero-shot performance reported above sets the upper bound rather than guaranteeing the floor. Expanding the negative side with primate calls, heavy rainfall, cicada choruses, and tropical anthropogenic noise from underrepresented regions is the most direct path to lower field-time false positives.

\paragraph{On-AudioMoth measurement is projected, not measured.}
The battery-life numbers in Table~\ref{tab:battery_projection} are projections from \citet{benhammadi2026battery}'s electrical coefficients (their Mel detector shares DrongoNet-Micro's CMSIS-DSP mel front-end). They are not direct on-AudioMoth INA219 measurements; the on-device measurements in this paper come from the Portenta H7 (Cortex-M7, Section~\ref{sec:deployment}), which has roughly an order of magnitude higher current envelope. Porting the DrongoNet-Micro firmware to the AudioMoth EFM32 Wonder Gecko and re-running the Benhammadi protocol is the natural validation step and is stated as immediate future work.

\paragraph{Field validation.}
The recall, FPR, storage, and projected-battery numbers are all derived from curated test splits and a single yellowhammer-survey energy model. Confirming them in the wild would require multi-week deployments in tropical forest, covering uncurated recordings, variable weather, multi-month reliability, and tropical-specific vocal-rate distributions. That validation lies beyond this paper's algorithmic scope.

\paragraph{Gatekeeper-only outputs.}
The model deliberately omits species-level outputs. This is by design (the binary head fits the MCU budget; a downstream classifier handles taxa), but it means DrongoNet alone cannot answer ecological questions about community composition---only presence-absence and segment-level activity rates.

\paragraph{Future work.}
Three directions follow directly from the limitations above. (i) An on-AudioMoth firmware port and INA219 validation of Table~\ref{tab:battery_projection}---the most direct way to close the projection-vs-measurement gap and to publish a head-to-head update against \citet{benhammadi2026battery}'s Table~4. (ii) Like-for-like comparison against BirdVoxDetect~\citep{lostanlen2024birdvoxdetect} and fine-tuned BirdNET on SEABAD, and cross-regional generalisation to Amazonian, African, and Oceanian rainforests. (iii) Integration with species-level classifiers in cascaded pipelines that exploit the gatekeeper-then-classifier structure---using Micro as the wake-up trigger and a larger off-device or on-SBC model for taxonomy, so the energy cost of the species head is gated by Micro's high-recall, high-precision-via-threshold-tuning binary decision.

\section{Conclusion}
\label{sec:conclusion}
\label{sec:conc}

DrongoNet is a family of ultra-lightweight CNNs for tropical bird presence detection on embedded hardware. A four-phase ablation identified the design choices that matter at MCU scale: global average pooling for parameter efficiency, focal loss for improved low-threshold calibration, and a lightweight learnable spectral weighting layer that replaces batch normalisation where BN becomes unstable after quantisation. The same study also showed that depthwise-separable convolutions provide no benefit at $n_\mathrm{mels}=16$, despite their widespread use in compact CNNs.

The three deployed variants (Table~\ref{tab:model_variants}) form a Pareto-efficient size--accuracy trade-off. \textbf{Nano} (5.09\,kB, $0.9727$ AUC) targets severely memory-constrained MCUs. \textbf{Micro} (6.26\,kB, $0.9810$ AUC, $98.3\%$ recall at $\tau=0.35$) is the recommended configuration for AudioMoth-class devices. \textbf{Edge} (33.06\,kB, $0.9991$ AUC, $99.0\%$ recall at $\tau=0.425$) is intended for Linux-based embedded systems. Compared with an equal-architecture retrained TinyChirp baseline (0.9815 AUC), Edge improves AUC by 1.8 percentage points at the same parameter budget, while Micro achieves comparable discrimination within 0.1 percentage points using $28\times$ fewer parameters. Across all three variants, full-INT8 quantisation reduces AUC by less than 0.12\%.

These results address a practical limitation of autonomous bioacoustic monitoring. In tropical deployments, most recording time is occupied by wind, rain, insects, and other non-target sounds, while existing detectors often miss a substantial proportion of bird vocalisations in Southeast Asian soundscapes~\citep{funosas2026global}. At $\tau=0.35$, DrongoNet-Micro detects eight percentage points more bird vocalisations than the Goertzel trigger used in current AudioMoth firmware while extending the estimated recording duration of a 32\,GB memory card from approximately 28 to 45 days. Achieving this performance with a 6.26\,kB model demonstrates that effective bird-presence detection can be deployed within the memory constraints of widely used field hardware.

Together with the SEABAD corpus~\citep{zabidi2026seabad}, DrongoNet provides a practical foundation for tropical bird-presence detection and supports a gatekeeper--classifier architecture in which a lightweight detector filters recordings before species-level classification. This approach reduces storage and energy consumption while preserving high recall, making long-term autonomous monitoring more practical for deployments in Southeast Asia and other underrepresented regions.

Future work will examine how much further the detector can be compressed before recall degrades to unacceptable levels, as well as validating long-term performance under continuous field deployment. These directions will help establish the practical limits of kilobyte-scale neural detectors for autonomous bioacoustic monitoring.

\section*{Data and Code Availability}

The SEABAD dataset is publicly available on Zenodo. All model architectures, training code, and evaluation protocols are released as open-source software:

\begin{itemize}
\item \textbf{Dataset}: \url{https://zenodo.org/records/18290494}
\item \textbf{Curation pipeline}: \url{https://github.com/mun3im/seabad}
\item \textbf{Model code}: \url{https://github.com/mun3im/drongonet}
\end{itemize}

The repository includes trained model weights (TFLite and TensorFlow formats), training scripts, evaluation code, and INT8 quantisation implementations, released under an open-source license.

\section*{Funding}

This research received no specific grant from any funding agency in the public, commercial, or not-for-profit sectors.

\section*{CRediT authorship contribution statement}

\textbf{Muhammad Mun'im Ahmad Zabidi:} Conceptualization, Methodology, Software, Validation, Formal analysis, Investigation, Data Curation, Writing -- Original Draft, Writing -- Review \& Editing, Visualization.
\textbf{Mohd Yamani Idna Idris:} Supervision, Writing -- Review \& Editing.
\textbf{Norisma Idris:} Supervision, Writing -- Review \& Editing.

\section*{Acknowledgments}

The authors thank Ben Ngui Yu Wei for assistance flashing the DrongoNet firmware to the Portenta H7 and extracting the on-device latency, memory, and power logs used in Section~\ref{sec:deployment}.

\section*{Declaration of Competing Interest}

The authors declare that they have no known competing financial interests or personal relationships that could have appeared to influence the work reported in this paper.

\bibliographystyle{abbrvnat}
\bibliography{references}

\appendix

\section{SEABAD Dataset: Detailed Curation Methodology}
\label{app:dataset}

\paragraph{Curation Pipeline.}
SEABAD's construction follows a six-stage pipeline: (1) Xeno-Canto acquisition from Malaysia and neighbouring Southeast Asian countries, (2) format unification (FLAC, 16\,kHz mono) with peak normalisation and soft-clip mitigation, (3) FAISS-based acoustic deduplication over mel-spectrogram embeddings, (4) fixed-length segment extraction with RMS filtering and temporal diversity scoring, (5) diversity-aware species balancing via acoustic clustering and salience-weighted stratified sampling, and (6) negative-sample curation from multiple complementary corpora.

\paragraph{Train/Test Split and Leakage Prevention.}
Recording-level leakage is structurally precluded: the pipeline extracts exactly one 3-second clip per Xeno-Canto recording, so no recording spans partitions. Negative samples come from distinct corpora (DCASE, ESC-50, FSC-22, DataSEC), each assigned to a single partition. FAISS deduplication is applied before splitting \citep{cramer2019look}. All seeds share an identical split; AUC variance reflects model stochasticity only.

\paragraph{Dataset Statistics.}
25,000 positive samples (1,677 species, mean 14.9/species, Gini 0.519) and 25,000 negative samples (17,685 from DCASE; 7,315 from ESC-50, FSC-22, DataSEC). 92.1\% of positives are rated Xeno-Canto class A/B.

\paragraph{Embedded Compatibility.}
3-second / 16\,kHz clips (48,000 samples) match typical ARU buffer sizes and cover avian vocalisations up to the 8\,kHz Nyquist limit at $3\times$ lower memory cost than 48\,kHz audio.

\end{document}